\documentclass[english,nofootinbib]{article}
\usepackage{ae,aecompl}
\usepackage[T1]{fontenc}
\usepackage[latin9]{inputenc}
\usepackage[letterpaper]{geometry}
\usepackage{prettyref}
\usepackage{amsmath}
\usepackage{setspace}
\usepackage{amssymb}

\usepackage{ae}
\usepackage{aecompl}

\usepackage[sf,bf,small]{titlesec}

\newcommand{\ket}[1]{\left | #1 \right >}

\makeatother

\usepackage{babel}

\usepackage{babel}

\begin{document}
\begin{flushright}
BROWN-HET 1554\\
arXiv:0809.2645
\par\end{flushright}

~

\begin{doublespace}
\begin{center}
\textsf{\textbf{\Large Matrix Reduction and the $\mathfrak{su}\left(2|2\right)$Superalgebra
in AdS/CFT}}
\par\end{center}{\Large \par}
\end{doublespace}

~

\begin{onehalfspace}
\begin{center}
In\^{e}s V. Aniceto\\
\emph{Brown University, Providence RI, 02912 USA.}\\
\emph{nes@het.brown.edu}\\
~\\
 28th April 2009
\par\end{center}
\end{onehalfspace}
\begin{abstract}
We study the supersymmetry generators $Q,\, S$ on the 1-loop vectorless
sector of $\mathcal{N}=4$ super Yang-Mills, by reduction to the plane-wave
matrix model. Using a coherent basis in the $\mathfrak{su}\left(2|2\right)$
sector, a comparison with the algebra given by Beisert in nlin/0610017
is presented, and some parameters (up to one-loop) are determined.
We make a final comparison of these supercharges with the results
that can be obtained from the string action by working in the light-cone-gauge
and discretizing the string.
\end{abstract}

\section{Introduction}

The gauge/string duality has been the object of study for more than
a decade by means of the AdS/CFT correspondence, \cite{Maldacena:1997re,Gubser:1998bc,Witten:1998qj}
between IIB superstrings on $AdS_{5}\times S^{5}$ and $\mathcal{N}=4$
$U\left(N_{c}\right)$ super Yang-Mills theory in four dimensions.
But while the results calculated from the gauge theory are perturbative
in 't Hooft coupling $\lambda=g_{YM}^{2}N_{c}$, the calculations
on the string side are valid for strong coupling $\lambda$.

This strong/weak property of the duality limited its study to operators/states
in sectors protected by supersymmetry, as these would receive no quantum
corrections. But a heuristic comparison of the algebraic structures
in the weak/strong coupling limits was possible by taking the plane-wave
limit, or BMN limit \cite{Berenstein:2002jq}. On the gauge theory
side, the BMN limit is taken by considering single trace operators,
i.e. $N_{c}$ very large, with large $R$-charge of $\mathfrak{so}\left(6\right)$
$J\sim\sqrt{N_{c}}$ and conformal dimension $\Delta$ , keeping $\Delta-J$
finite. These operators consist of a chiral primary (trace of a large
number of a complex field) with some impurities (other complex fields,
bosonic or fermionic). Even though the 't Hooft coupling $\lambda=g_{YM}^{2}N_{c}$
is very large, one can use perturbation theory provided some effective
coupling $\lambda'=g_{YM}^{2}N_{c}/J^{2}\sim g_{YM}^{2}$ is kept
fixed and small.

On the string side we start from the Green-Schwarz action on the $AdS_{5}\times S^{5}$
\cite{Metsaev:1998it}, with $J\sim\sqrt{N_{c}}$ now being the angular
momentum in one of the directions of $S^{5}$. We also take the energy
$E$ (generator of time translations in $AdS_{5}$) to be large, obeying
$E-J$ finite, thus originating point-like closed strings with large
angular momentum in $S^{5}$. In light-cone gauge, the quantity $E-J$
is just the light-cone hamiltonian, and the light-cone momentum $P_{+}=E+J$
is very large. In this case, there is an effective coupling just $\tilde{\lambda}=4\lambda/P_{+}^{2}$,
which is equivalent to $\lambda'$ in the limit $J\rightarrow\infty$
($P_{+}/2\rightarrow J$).This limit allowed a direct comparison of
the dilatation operator in SYM (anomalous dimensions of operators
in the conformal field theory) to the energies ($E-J$) of point like
semiclassical string oscillations in the plane-wave geometry.

The algebra of symmetries, $\mathfrak{psu}\left(2,2|4\right)$ is
central in AdS$_{5}$/CFT$_{4}$ correspondence, as both the gauge
theory and its string theory dual have the same underlying supersymmetry
algebra. The two dimensional sigma-model which gives us the perturbative
string theory in $AdS_{5}\times S^{5}$ \cite{Metsaev:1998it} has
a manifest global symmetry under $\mbox{PSU}\left(2,2|4\right)$ \cite{Gunaydin:1984fk,Kim:1985ez},
which is the same group of internal and space-time symmetries of the
$\mathcal{N}=4$ SYM (see \cite{Beisert:2004ry} and references therein).

In particular, one can use the algebra to compare the scattering of
particles in the duality. For large 't Hooft coupling the scattering
is best described by string theory, but for small 't Hooft coupling
the spin-chain description is more adequate. It was shown by Beisert
that the non-pertubative S-matrix is almost completely determined
by the centrally extended algebra $\mathfrak{su}\left(2|2\right)\oplus\mathfrak{su}\left(2|2\right)$
\cite{Beisert:2005tm,Beisert:2006qh}, up to an overall dressing phase
(determined by a crossing symmetry restriction \cite{Janik:2006dc,Beisert:2006ib,Beisert:2006ez,Beisert:2006zy}).
Each of these centrally extended algebras $\mathfrak{su}\left(2|2\right)$
has the following structure: bosonic (kinematical) generators $\mathbf{R}_{\, b}^{a},\mathbf{L}_{\,\beta}^{\alpha}$,
corresponding to the rotation generators of the bosonic subalgebra
$\mathfrak{su}\left(2\right)\oplus\mathfrak{su}\left(2\right)$; fermionic
(dynamical) supersymmetry generators $\mathbf{Q}_{\, a}^{\alpha},\,\mathbf{Q}_{\;\beta}^{\dagger b}$;
and three central charges $\mathbf{H},\,\mathbf{C},\,\mathbf{C}^{\dagger}$(hamiltonian,
generator of space translations and of boosts).%
\footnote{Note that $\left(\mathbf{Q}_{\, a}^{\alpha}\right)^{\dagger}=\mathbf{Q}_{\;\alpha}^{\dagger a}$
and the same relation holds to the central elements $\mathbf{C}$
and $\mathbf{C}^{\dagger}$.%
} Their commutation relations are:\begin{eqnarray}
\left[\mathbf{L}_{\,\beta}^{\alpha},\mathbf{J}^{\gamma}\right] & = & \delta_{\beta}^{\gamma}\mathbf{J}^{\alpha}-\frac{1}{2}\delta_{\beta}^{\alpha}\mathbf{J}^{\gamma}\:,\qquad\left[\mathbf{R}_{\, b}^{a},\mathbf{J}^{c}\right]=\delta_{b}^{c}\mathbf{J}^{a}-\frac{1}{2}\delta_{b}^{a}\mathbf{J}^{c}\,,\nonumber \\
\left[\mathbf{L}_{\,\beta}^{\alpha},\mathbf{J}_{\gamma}\right] & = & -\delta_{\gamma}^{\alpha}\mathbf{J}_{\beta}+\frac{1}{2}\delta_{\beta}^{\alpha}\mathbf{J}_{\gamma}\:,\qquad\left[\mathbf{R}_{\, b}^{a},\mathbf{J}_{c}\right]=-\delta_{c}^{a}\mathbf{J}_{b}+\frac{1}{2}\delta_{b}^{a}\mathbf{J}_{c}\,,\nonumber \\
\left\{ \mathbf{Q}_{\, a}^{\alpha},\mathbf{Q}_{\;\beta}^{\dagger b}\right\}  & = & \delta_{a}^{b}\mathbf{L}_{\,\beta}^{\alpha}+\delta_{\beta}^{\alpha}\mathbf{R}_{\, a}^{b}+\frac{1}{2}\delta_{a}^{b}\delta_{\beta}^{\alpha}\mathbf{H},\label{Commutation relations of symmetry generators}\\
\left\{ \mathbf{Q}_{\, a}^{\alpha},\mathbf{Q}_{\, b}^{\beta}\right\}  & = & \epsilon^{\alpha\beta}\epsilon_{ab}\mathbf{C}\:,\qquad\left\{ \mathbf{Q}_{\;\alpha}^{\dagger a},\mathbf{Q}_{\;\beta}^{\dagger b}\right\} =\epsilon^{ab}\epsilon_{\alpha\beta}\mathbf{C}^{\dagger}\,.\nonumber \end{eqnarray}
 In the above expressions, $\mathbf{J}^{M}$ (where $M\in\left\{ a,\alpha\right\} $,
$a$ being bosonic indices and $\alpha$ being the fermionic ones),
is any element of the Lie algebra. From the elements of the algebra,
the dilatation operator (or central charge hamiltonian) has been studied
in detail (see \cite{Beisert:2004ry} and references therein).

Much has been done on the study of sectors of this superconformal
algebra on the string side \cite{Arutyunov:2005hd,Arutyunov:2006yd,Frolov:2006cc,Klose:2006zd,Arutyunov:2006ak}.
On the gauge side, Beisert has perturbatively studied and determined
the action of the generators of the superalgebra $\mathfrak{su}\left(2|2\right)$
up to two loops, by first restricting to the subalgebra $\mathfrak{su}\left(2|3\right)$
whose fundamental representation consists of three complex scalars
and two complex fermions \cite{Beisert:2003ys}, and finally considering
an infinite chain of one of the scalar operators \cite{Beisert:2005tm}.
Using Bethe Ansatz techniques it was later conjectured an all loop
result in this sector for the action of the algebra generators \cite{Beisert:2006qh}.

In this work, we present the SUSY algebra in terms of a matrix model
reduction of Yang-Mills Theory in the large $N$ limit. The matrix
model has played a very useful role in large $N$ theories. In fact,
the $\frac{1}{2}$ Bogomol'nyi-Prasad-Sommerfield (BPS) sector of
$\mathcal{N}=4$ SYM is completely described in terms of a complex
matrix model \cite{Corley:2001zk,Berenstein:2004kk,Lin:2004nb,Donos:2005vm,Cremonini:2007kh},
and the $\frac{1}{4}$ BPS generalization is also of great interest
(work in progress). Presently, the interest is in the detailed construction
and comparison of supercharges and their commutation relations both
on the Yang-Mills and on the string side. We will demonstrate that
the algebra given by Beisert in \cite{Beisert:2006qh} (at least at
one loop) is correctly reproduced from the reduced Matrix model point
of view.

In \cite{Kim:2003rz} it was seen that the plane-wave matrix theory
\cite{Berenstein:2002jq,Kim:2002if} arises when compactifying $\mathcal{N}=4$
SYM in $\mathbb{R}\times S^{3}$ followed by a consistent truncation
in order to keep only the lowest Kaluza-Klein modes (see also \cite{Okuyama:2002zn,Ishiki:2006rt}).
These modes have masses proportional to a mass parameter, given by
$\left(\frac{m}{3}\right)^{3}=\frac{32\pi^{2}}{g_{YM}^{2}}$. This
theory was shown (in some sectors) to still be integrable up to four-loops
\cite{Klose:2003qc,Fischbacher:2004iu}. Study of this model is simpler
than the full $\mathcal{N}=4$ SYM, and can be found in \prettyref{sec:N=00003D00003D4-sym-on-RxS3-review}.
In this section we present a detailed study of the supercharges $Q$
and $S$, following the approach of \cite{Kim:2003rz}. In \prettyref{sec:su(2|3) sector and restr to su(2|2)}
we restrict the action of the generators of the algebra to a subsector
$\mathfrak{su}\left(2|2\right)$. The results presented in this paper
are one-loop, and we compare our results with the non-local generators
presented in \cite{Beisert:2006qh}, evaluating some of the parameters
defining these generators.

Some methods have been employed in the gauge theory side that allowed
a comparison of the hamiltonian to string theory equivalent algebra
generator. Such methods include the use of coherent states, \cite{Kruczenski:2003gt,Kruczenski:2004kw,Mikhailov:2004xw}
collective field theory and string field theory \cite{deMelloKoch:2002nq,deMelloKoch:2003pv}.
In this framework one can compare a discrete (first quantized) version
of the supercharges on the string side with the oscillator expansion
of the charges in SYM, in the BMN limit.

In \prettyref{sec:su(2|3) sector and restr to su(2|2)} we use a coherent
state basis to write the supercharges, which are then shown, in \prettyref{sec:SUSY-generators-in AdSxS},
to have the same structure as the first quantized version of the algebra
generators determined from the string action. In \prettyref{sec:Conclusions}
we present a summary of the results, and some future paths of investigation.

\section{$\mathcal{N}=4$ $\mathrm{SYM}$ on $\mathbb{R}\times S^{3}$: A
review\label{sec:N=00003D00003D4-sym-on-RxS3-review}}

In this first section, we summarize the method of finding the supercharges
of $\mathfrak{su}\left(2,2|4\right)$ up to 1-loop, as can be found
in \cite{Okuyama:2002zn,Kim:2003rz}.%
\footnote{We will be following the notation of \cite{Kim:2003rz}, in which
a different basis for the $\gamma-$matrices is used. The same procedure
could be done by following \cite{Okuyama:2002zn} choice of basis.%
}

The action for $\mathcal{N}=4$ $\mathrm{SYM}$ in four dimensions
can be obtained from dimensional reduction of the $\mathcal{N}=1$
10 dimensional $\mathrm{SYM}$ on a $6-$torus. Using the notation
where the $D=10$ Dirac matrices split into $\mathrm{SO}\left(1,3\right)\times\mathrm{SO}\left(6\right)$,
the action becomes:\begin{eqnarray*}
S & = & \frac{2}{g_{YM}^{2}}\int d^{4}x\sqrt{\left|g\right|}\mathrm{Tr}\left\{ -\frac{1}{4}F^{\mu\nu}F_{\mu\nu}-\frac{1}{2}D^{\mu}\phi_{i}D_{\mu}\phi_{i}-\frac{\mathcal{R}}{12}\phi_{i}^{2}+\frac{1}{4}\left[\phi_{i},\phi_{j}\right]^{2}-2i\lambda_{A}^{\dagger}\sigma^{\mu}D_{\mu}\lambda^{A}+\right.\\
 &  & \qquad\qquad\qquad\qquad\qquad\left.+\left(\rho_{i}\right)^{AB}\lambda_{A}^{\dagger}i\sigma^{2}\left[\phi_{i},\lambda_{B}^{*}\right]-\left(\rho_{i}^{\dagger}\right)_{AB}\left(\lambda^{A}\right)^{T}i\sigma^{2}\left[\phi^{i},\lambda^{B}\right]\right\} \,.\end{eqnarray*}

We have a vector field $A_{\mu}$, six real scalars $\phi_{i}$ and
four Weyl spinors $\lambda_{\alpha A}$ (all in the adjoint representation
of the gauge group). The six scalars transform in a $\mathbf{6}$
of the $R-$symmetry group $SO\left(6\right)\equiv SU\left(4\right)_{R}$,
while the spinors transform in a $\mathbf{4}$. Coordinate indices
are $x^{\mu}=\left(t,x^{a}\right)$, $\mu=0,...,3$, with the spacial
coordinates having (curved) indices $a=1,2,3$. The metric is given
by \[
ds^{2}=-dt^{2}+R^{2}\left(d\theta^{2}+\sin^{2}\theta d\psi^{2}+\sin^{2}\theta\sin^{2}\psi d\chi^{2}\right),\]
 where $R$ is the radius of $S^{3}$, and $\mathcal{R}=\frac{6}{R^{2}}$
is the Ricci scalar.

\subsubsection*{Some Notation}

From this point on we will be considering $\sigma^{\mu}\equiv\left(\mathbf{1},\sigma^{a}\right)$
and $\overline{\sigma}^{\mu}=\left(-\mathbf{1},\sigma^{a}\right)$,
where the $\sigma^{a}$ are the usual Pauli matrices pulled back to
$S^{3}$. Also, $\rho_{i}^{AB}\equiv\sigma_{i}^{AB}$ are the Clebsch-Gordan
coefficients of $SU\left(4\right)$ that relate two $\mathbf{4}$
irreducible representations (irreps) with one $\mathbf{6}$. These
coefficients have several properties, in particular $\rho_{i}^{AB}=\frac{1}{2}\varepsilon^{ABCD}\left(\rho_{i}^{\dagger}\right)_{CD}$,
and allow us to write \[
\phi_{i}=\frac{1}{2}\rho_{i}^{AB}\Phi_{AB}=\frac{1}{2}\left(\rho_{i}^{\dagger}\right)_{AB}\Phi^{AB}.\]

Finally, one comment about the Weyl spinors. We know that in $D=10$
we start from a 32-component complex spinor, and by imposing a Majorana-Weyl
condition, obtain a 16-component (after fixing the $\kappa$-symmetry)
spinor $L$. This spinor can be written in terms of Weyl spinors as\[
L=\left(\begin{array}{c}
\lambda^{\alpha A}\\
i\left(\sigma^{2}\right)^{\alpha\beta}\lambda_{\beta A}^{*}\end{array}\right),\]
 with $\alpha=1,2$ and $A=1,2,3,4$. The $\lambda^{\alpha A}$ are
four 2-component Weyl spinors.%
\footnote{ In the basis used in \cite{Okuyama:2002zn}, the separation of $L$
into $L=\left(L_{+}\; L_{-}\right)^{T}$ becomes a separation into
$SU\left(2\right)_{L}\times SU\left(2\right)_{R}$, for which one
uses dotted/undotted indices $\dot{\alpha},\alpha$. In the basis
used in \cite{Kim:2003rz} this separation is not obvious.%
}

\subsection{SUSY transformations and corresponding charges}

The SUSY transformations are given by:\begin{eqnarray*}
\delta_{\eta}A_{\mu} & = & 2i\left(\lambda_{A}^{\dagger}\sigma_{\mu}\eta^{A}-\eta_{A}^{\dagger}\sigma_{\mu}\lambda^{A}\right),\\
\delta_{\eta}\Phi^{AB} & = & 2i\left(-\lambda_{E}^{\dagger}i\sigma^{2}\varepsilon^{ABEF}\eta_{F}^{*}-\left(\lambda^{A}\right)^{T}i\sigma^{2}\eta^{B}-\left(\lambda^{B}\right)^{T}i\sigma^{2}\eta^{A}\right),\\
\delta_{\eta}\lambda^{A} & = & \frac{1}{2}F_{\mu\nu}\sigma^{\mu\nu}\eta^{A}+2D_{\mu}\Phi^{AB}\overline{\sigma}^{\mu}i\sigma^{2}\eta_{B}^{*}+\Phi^{AB}\overline{\sigma}^{\mu}i\sigma^{2}\nabla_{\mu}\eta_{B}^{*}-2i\left[\Phi^{AC},\Phi_{CB}\right]\eta^{B}.\end{eqnarray*}

We want to build the Noether charge $Q\eta$ . To do so we need to
take into consideration the pairs of canonical variables. From the
action, we have the following (anti-)commutation relations:\begin{eqnarray*}
\left[F_{0\mu},A^{\nu}\right] & = & \delta_{\mu}^{\nu},\\
\left[D_{0}\phi_{i},\phi_{j}\right] & = & \delta_{ij}\quad\Rightarrow\left[D_{0}\Phi_{AB},\Phi^{CD}\right]=\frac{1}{2}\left(\delta_{A}^{D}\delta_{B}^{C}-\delta_{A}^{C}\delta_{B}^{D}\right),\\
\left\{ -i\left(\lambda_{A}^{\dagger}\sigma^{0}\right)_{\alpha},\lambda^{B\beta}\right\}  & = & \delta_{\alpha}^{\beta}\delta_{A}^{B}.\end{eqnarray*}

Also one has to take into consideration that $\eta^{\alpha A}$ are
Killing spinors, which in $\mathbb{R}\times S^{3}$ obey the equation
$\nabla_{\mu}\eta=\pm\frac{i}{2R}\sigma_{\mu}\eta,$ and so will give
us two solutions $\eta_{\pm}$. We will then obtain two charges $Q\equiv Q_{L}$
and $\overline{Q}\equiv Q_{R}$, corresponding to $\eta_{+}$ and
$\eta_{-}$ respectively.

The fermionic Noether charges are thus%
\footnote{For comparison purposes, one could also write this charge, in the
$SU\left(2\right)_{L}\times SU\left(2\right)_{R}$ formalism, as\[
Q_{\epsilon}=\frac{2}{g_{YM}^{2}}\int_{s^{3}}\mathrm{Tr}\left\{ i\overline{\lambda}_{A}^{\dot{\alpha}}\overline{\sigma}_{\dot{\alpha}\alpha}^{0}\delta_{\epsilon}\lambda^{\alpha A}+i\overline{\lambda}_{\alpha}^{A}\left(\sigma^{0}\right)^{\alpha\dot{\alpha}}\delta_{\epsilon}\lambda_{\dot{\alpha}A}\right\} .\]
}\[
Q\eta=\frac{2}{g_{YM}^{2}}\int_{S^{3}}d\Omega\,\mathrm{Tr}\left\{ -2i\lambda_{A}^{\dagger}\sigma^{0}\delta_{\eta}\lambda^{A}-2i\left(\lambda^{A}\right)^{T}\sigma^{0}\delta_{\eta}\lambda_{A}^{*}\right\} .\]

For the purposes of this paper, we will simplify the calculations
by setting the vector field to zero (we will be looking only at the
sector of scalars and spinors). This truncation is consistent with
the one-loop calculation we will be performing.

The non-vector sector of the charges $Q\eta$ is given by:{\small \begin{eqnarray}
Q\eta & =- & \frac{2}{g_{YM}^{2}}\int_{S^{3}}d\Omega\,\mathrm{Tr}\left\{ 2i\lambda_{A}^{\dagger}\left(2\Pi^{AB}\sigma^{0}i\sigma^{2}\eta_{B}^{*}+2\nabla_{a}\Phi^{AB}\overline{\sigma}^{a}i\sigma^{2}\eta_{B}^{*}+\Phi^{AB}\overline{\sigma}^{\mu}i\sigma^{2}\nabla_{\mu}\eta_{B}^{*}-2i\left[\Phi^{AC},\Phi_{CB}\right]\eta^{B}\right)+\right.\nonumber \\
 &  & \left.2i\left(\lambda^{A}\right)^{T}\left(-2\Pi_{AB}\left(\overline{\sigma}^{0}\right)^{T}i\sigma^{2}\eta^{B}-2\nabla_{a}\Phi_{AB}\left(\overline{\sigma}^{a}\right)^{T}i\sigma^{2}\eta^{B}-\Phi_{AB}\left(\overline{\sigma}^{\mu}\right)^{T}i\sigma^{2}\nabla_{\mu}\eta^{B}-2i\left[\Phi_{AC},\Phi^{CB}\right]\eta_{B}^{*}\right)\right\} ,\nonumber \\
 &  & \qquad\label{Supercharges, before integration}\end{eqnarray}
 }where $\Pi_{AB}$ is the momentum conjugate to the bosonic field
$\Phi^{AB}$.

We now have the an expression for the supercharges. The next step
is to evaluate it on $\mathbb{R}\times S^{3}$: we expand the four-dimensional
fields in terms of the spherical harmonics of $S^{3}$, and then perform
the integration of the sphere.

\subsection{Harmonic Expansion on $S^{3}$ and the Plane-Wave Limit}

Each field, defined by its spin, will have a decomposition in spherical
harmonics on $S^{3}$. These spherical harmonics can be labeled by
the irreducible representations $\left(m_{L},m_{R}\right)$ of the
isometry group $SO\left(4\right)\equiv SU\left(2\right)_{L}\otimes SU\left(2\right)_{R}$.
As such, we have:
\begin{itemize}
\item Spin $0$: We have scalar spherical harmonics $Y_{\left(0\right)}^{k\, I}$,
in the irrep $\left(k+1,k+1\right)$. Their mass will be $\left(k+1\right)/R$.
\item Spin $\frac{1}{2}$: In this case we'll use spinor spherical harmonics:
$Y_{\left(1/2\right)}^{k\, I\,+}$, in the irrep $\left(k+2,k+1\right)$;
$Y_{\left(1/2\right)}^{k\, I\,-}$, in the irrep $\left(k+1,k+2\right)$.
Both have mass $\left(k+3/2\right)/R$.
\end{itemize}
As usual, $k$ labels different irreducible representations, and $I$
enumerates the elements of a particular irrep ($I=1\cdots d$, where
$d$ is the dimension of the representation).

The expansions of the fields in the corresponding harmonics are:\begin{eqnarray*}
\phi_{i}\left(x^{\mu}\right) & = & \sum_{k=0}^{\infty}\sum_{I=1}^{\left(k+1\right)^{2}}\phi_{i}^{k\, I}\left(t\right)Y_{\left(0\right)}^{k\, I}\left(x^{a}\right),\\
\lambda_{\alpha}^{A}\left(x^{\mu}\right) & = & \sum_{k=0}^{\infty}\sum_{I=1}^{\left(k+1\right)\left(k+2\right)}\sum_{\pm}\lambda^{A,k\, I\,\pm}\left(t\right)Y_{\left(1/2\right)\,\alpha}^{k\, I\,\pm}\left(x^{a}\right).\end{eqnarray*}
 Note that spinor spherical harmonics are $2-$dimensional commuting
Weyl spinors. The Killing spinor $\eta^{A}$ (parameter of the superconformal
transformations) will have the same expansion as $\lambda^{A}$, with
coefficients $\eta^{A,k\, I\,\pm}\left(t\right)$.

\subsubsection*{Plane-Wave Limit }

We want to truncate the infinite tower of Kaluza-Klein modes to the
lowest supermultiplet \cite{Kim:2002if,Kim:2003rz}. One can then
climb up the various states (with increasing masses) by acting with
the two supercharges $Q_{L}=\left(\mathbf{2},\mathbf{1},\mathbf{\overline{4}}\right)$
and $Q_{R}=\left(\mathbf{1},\mathbf{2},\mathbf{4}\right)$, where
the numbers correspond to representations of $SU\left(2\right)_{L}\otimes SU\left(2\right)_{R}\otimes SU\left(4\right)$.
Focusing on the zero modes of the Kaluza-Klein tower we find 6 scalar
spherical harmonics, constant on $S^{3}$, and 4 lowest spinor spherical
harmonics $S_{\alpha}^{\hat{\alpha}\,\pm}$, in irrep $\left(2,1\right)\oplus\left(1,2\right)$
of $SU\left(2\right)_{L}\otimes SU\left(2\right)_{R}$ (the hatted
index refers to the degeneracy of the solution), solutions to the
the killing spinor equation for a Weyl spinor.

The fields with only these zero modes become:\begin{eqnarray*}
\phi_{i}\left(x^{\mu}\right) & = & X_{i}\left(t\right),\\
\lambda_{\alpha}^{A}\left(x^{\mu}\right) & = & \sum_{\hat{\alpha}=1}^{2}\left(\theta_{\hat{\alpha}}^{A\,+}\left(t\right)S_{\alpha}^{\hat{\alpha}\,+}\left(x^{a}\right)+\theta_{\hat{\alpha}}^{A\,-}\left(t\right)S_{\alpha}^{\hat{\alpha}\,-}\left(x^{a}\right)\right).\end{eqnarray*}

If we restrict ourselves to half of the supercharges $Q_{L}$, then
together with the bosonic symmetries will generate the subalgebra
$\mathfrak{su}\left(2|4\right)$. The restriction to the $Q_{L}$
charges leads us to consider only the zero modes that are$SU\left(2\right)_{R}$
singlets. Then we keep all the lowest scalar harmonics, and only two
spinor harmonics $S_{\alpha}^{\hat{\alpha}\,+}$ (instead of the 4
if we included $S_{\alpha}^{\hat{\alpha}\,-}$). The conjugate momenta
$\pi_{i}$ will have the same expansion as its conjugate field $\phi_{i}$,
that is $\pi_{i}\left(x^{\mu}\right)=\Pi_{i}\left(t\right)$.

Now we can proceed to the actual integration on the supercharges.
Going back to (\ref{Supercharges, before integration}), we find that:%
\footnote{In order to obtain the supercharges integrated over $S^{3}$, we used
the properties of the spherical harmonics, as well as other properties
of the Pauli matrices. These properties can be found in \cite{Okuyama:2002zn,Kim:2002if,Kim:2003rz},
and include $\overline{\sigma}^{\mu}i\sigma^{2}\sigma_{\mu}^{T}=\left(\overline{\sigma}^{\mu}\right)^{T}i\sigma^{2}\sigma_{\mu}=-2i\sigma^{2}$.
In the same references one can find the expansion of spin 1 vector
fields. We also used an identification between the radius of the sphere
$R$ and the Yang-Mills coupling constant $g_{YM}$ such that $4\pi^{2}R^{3}/g_{YM}^{2}\rightarrow1$.
This prefactor shows up when obtaining the action of the plane-wave
matrix theory action from $\mathcal{N}=4$ SYM action, and would also
appear in the charges.%
}\begin{eqnarray*}
Q_{L} & = & Q\eta^{+}=\mathrm{Tr}\left\{ \left(\frac{1}{R}\, X^{AB}+2i\Pi^{AB}\right)\theta_{A}^{+\dagger}i\sigma^{2}\eta_{B}^{+*}-\sqrt{2}\left[X_{AC},X^{CB}\right]\theta_{\hat{\alpha}}^{+A}\varepsilon^{\hat{\alpha}\hat{\beta}}\eta_{B\hat{\beta}}^{+*}\right.\\
 &  & \qquad\qquad\qquad\left.+\left(\frac{1}{R}\, X_{AB}-2i\Pi_{AB}\right)\left(\theta^{+A}\right)^{T}i\sigma^{2}\eta^{+B}-\sqrt{2}\left[X^{AC},X_{CB}\right]\theta_{A\hat{\alpha}}^{+\dagger}\varepsilon^{\hat{\alpha}\hat{\beta}}\eta_{\hat{\beta}}^{+B}\right\} \\
 & = & Q_{+}\eta+S_{+}\eta^{*}.\end{eqnarray*}

The final expression for the supercharges is %
\footnote{Note that in our choice of basis the relation $S=Q^{\dagger}$ is
not manifest. %
}

\begin{eqnarray}
Q_{A}^{\hat{\alpha}} & = & \mathrm{Tr}\left\{ -\theta^{B\hat{\alpha}}\left(\frac{1}{R}X_{BA}-2i\Pi_{BA}\right)-\sqrt{2}\varepsilon^{\hat{\beta}\hat{\alpha}}\theta_{B\hat{\beta}}^{\dagger}\left[X^{BC},X_{CA}\right]\right\} ,\nonumber \\
S^{A\hat{\alpha}} & = & \mathrm{Tr}\left\{ \theta_{B\hat{\beta}}^{\dagger}\left(\frac{1}{R}X^{BA}+2i\Pi^{BA}\right)-\sqrt{2}\left[X_{BC},X^{CA}\right]\theta_{\hat{\beta}}^{B}\right\} \varepsilon^{\hat{\beta}\hat{\alpha}}.\end{eqnarray}

\section{The $\mathfrak{su}\left(2|3\right)$ subsector and its restriction
to the $\mathfrak{su}\left(2|2\right)$\label{sec:su(2|3) sector and restr to su(2|2)}}

We'll continue by studying the sector $\mathfrak{su}\left(2|3\right)$,
as in \cite{Beisert:2003ys}. For that we reduce our fields as follows:\[
\theta^{\alpha}\equiv\theta^{4\hat{\alpha}},\;\phi^{a}\equiv X^{a4},\:\alpha=1,2;\: a=1,2,3.\]
 By construction we have $\overline{\phi^{a}}\equiv\phi_{a}$, and
$\pi_{a}=\Pi_{4a}$, as well as $X^{BC}=\frac{1}{2}\varepsilon^{BCAD}X_{AD}$.
The supercharges restricted to this sector can then be written as:\begin{eqnarray}
Q_{\, a}^{\alpha} & = & \mathrm{Tr}\left\{ -\theta^{4\hat{\alpha}}\left(\frac{1}{R}X_{4a}-2i\Pi_{4a}\right)-\sqrt{2}\theta_{4\hat{\beta}}^{\dagger}\varepsilon^{\hat{\beta}\hat{\alpha}}\left[X^{4C},X_{Ca}\right]\right\} \nonumber \\
 & = & \mathrm{Tr}\left\{ \theta^{\alpha}\left(\frac{1}{R}\overline{\phi}_{a}+2i\pi_{a}\right)-\sqrt{2}\theta_{\beta}^{\dagger}\varepsilon^{\alpha\beta}\varepsilon_{abc}\left[\phi^{c},\phi^{b}\right]\right\} ;\\
S^{a\,\alpha} & = & \mathrm{Tr}\left\{ \theta_{4\hat{\beta}}^{\dagger}\left(\frac{1}{R}X^{4a}+2i\Pi^{4a}\right)-\sqrt{2}\left[X_{4C},X^{Ca}\right]\theta_{\hat{\beta}}^{4}\right\} \varepsilon^{\hat{\beta}\hat{\alpha}}\nonumber \\
 & = & \mathrm{Tr}\left\{ \theta_{\beta}^{\dagger}\left(\frac{1}{R}\phi^{a}-2i\overline{\pi}^{a}\right)-\sqrt{2}\varepsilon^{abc}\left[\overline{\phi}_{c},\overline{\phi}_{b}\right]\theta^{\gamma}\varepsilon_{\gamma\beta}\right\} \varepsilon^{\beta\alpha}.\end{eqnarray}

In order to continue, we will need to rewrite the fields in terms
of creation/annihilation operators. First identify $\frac{1}{R}=\frac{m}{6}$,
i.e. exchange the parameter $R$ by a mass parameter $m$ \cite{Kim:2003rz}.
Then consider the expansion of the six scalars/momenta $X_{i},\,\Pi_{i}$:\[
\left\{ \begin{array}{l}
a_{i}=\sqrt{\frac{3}{m}}\left(i\Pi_{i}+\frac{m}{6}X_{i}\right),\\
a_{i}^{\dagger}=\sqrt{\frac{3}{m}}\left(-i\Pi_{i}+\frac{m}{6}X_{i}\right),\end{array}\right.\qquad\Rightarrow\quad\left\{ \begin{array}{l}
X_{i}=\sqrt{\frac{3}{m}}\left(a_{i}+a_{i}^{\dagger}\right),\\
\Pi_{i}=\frac{1}{2i}\sqrt{\frac{m}{3}}\left(a_{i}-a_{i}^{\dagger}\right).\end{array}\right.\]
 The bosons $X_{AB}$ are a combination of two real scalar fields
such that $X_{a4}=\frac{1}{2}\left(X_{a}+iX_{a+3}\right),\, a=1,2,3$.
If we now define the creation annihilation operators as $a^{a}\equiv a^{a}+ia^{a+3}$
and $b^{a\dagger}=a^{a\dagger}+ia^{a+3\dagger}$, with $a=1,2,3$,
we then have the following expansions for our (complex) fields:\begin{eqnarray}
\phi^{a} & \equiv X^{a4}= & \sqrt{\frac{3}{m}}\left(a^{a}+b^{\dagger a}\right)\,;\quad\pi_{a}\equiv\Pi_{4a}==\frac{1}{4i}\sqrt{\frac{m}{3}}\left(a_{a}^{\dagger}-b_{a}\right),\end{eqnarray}
 with equivalent expressions for fields $\overline{\phi}_{a}$ and
$\overline{\pi}^{a}$. Introducing also fermionic creation operators,
the fermions become \begin{equation}
\theta^{\dagger\alpha}=c^{\alpha}=\varepsilon^{\alpha\beta}c_{\beta}\;;\quad\theta^{\alpha}=c^{\dagger\alpha}.\end{equation}
 We will be interested in action of the charges on the subspace of
states that will only have excitations of $c^{\dagger}$ and $b^{\dagger}$,
so we will drop the oscillators $a,a^{\dagger}$ in the bosonic fields.
We find: \begin{eqnarray}
Q_{a}^{\,\alpha} & = & \mathrm{Tr}\left\{ \sqrt{\frac{m}{3}}\, c^{\dagger\alpha}b_{a}-\frac{3\sqrt{2}}{m}\varepsilon^{\alpha\beta}\varepsilon_{abc}\left[b^{\dagger c},b^{\dagger b}\right]c_{\beta}\right\} ,\nonumber \\
S_{\,\alpha}^{a} & = & \mathrm{Tr}\left\{ -\sqrt{\frac{m}{3}}\, b^{\dagger a}c_{\alpha}-\frac{3\sqrt{2}}{m}\varepsilon_{\alpha\beta}\varepsilon^{abc}c^{\dagger\beta}\left[b_{c},b_{b}\right]\right\} .\label{Suparcharges with oscillator}\end{eqnarray}
 As expected, these results are similar with the ones in \cite{Kim:2002if},
up to a change of basis for the gamma matrices.

\subsection{The $\mathfrak{su}\left(2|2\right)$ subsector: vacuum and excitations}

The study will focus on states that transform in the $\mathfrak{su}\left(2|3\right)$
sector and are single trace (gauge invariant) operators of the fields
(3 bosons and 2 fermions). This \emph{spin-chain} arises from the
large $N-$limit of the gauge theory. In this sector the action of
the algebra generators can be found in \cite{Beisert:2003ys}. Consider
now the vacuum as a long string of $Z\equiv\phi^{3}$ fields. In oscillator
notation, we have $Z=b^{3\dagger}$, and the vacuum state can be written
as:\[
\ket{0,J}\equiv\ket{Z^{J}}\equiv\frac{1}{\sqrt{J}N^{J/2}}\mathrm{Tr}\left(b^{3\dagger J}\right)\ket{0}.\]
 A generalization of this vacuum consists in an infinitely long string
of $Z$ fields (the asymptotic regime, $J\rightarrow\infty$), as
in \cite{Beisert:2005tm}. The excitations are now the other fields
of the $\mathfrak{su}\left(2|3\right)$ algebra, $\chi\in\left\{ \psi^{1},\psi^{2}|\phi^{1},\phi^{2}\right\} $,
which corresponds to the $\mathfrak{su}\left(2|2\right)$ subsector
of the algebra. The excitations can move through the chain on $Z's$
with some momentum $p$. Thus, in momentum space we can write \[
\chi=\sum_{n_{k}=1}^{N}e^{ip_{k}n_{k}}\chi\left(n_{k}\right)=\sum_{n_{k}=1}^{J}e^{ip_{k}n_{k}}\chi_{k}\equiv\chi\left(p_{k}\right),\]
 where $n$ denotes the position of the impurity/excitation $\chi$
on the vacuum string.

A general state with $K$ impurities can then be written as:\[
\ket{\chi_{1},...\chi_{K};J}=\sum_{n_{1},...,n_{K}=1}e^{ip_{1}n_{1}+\cdots+ip_{K}n_{K}}\ket{Z\cdots Z\chi_{1}Z\cdots\chi_{2}\cdots\chi_{K}...Z}.\]
 For an asymptotic state ( $J\rightarrow\infty$) we consider the
dilute gas approximation, where the positions $n_{1},\cdots,n_{k}$
of the impurities obey $n_{1}\ll n_{2}\ll\cdots\ll n_{K}$.

We should note that on-shell the physical states are cyclic (property
of the trace), and so we must have $\sum_{k=1}^{K}p_{k}=0$.

Now that we defined the sates that the supercharges will be acting
on, we can determine their action. The first step will be to check
what the charges do to just one excitation on the vacuum. Then one
can generalize to multi-excitation states of the $\mathfrak{su}\left(2|2\right)$
subsector of $\mathfrak{su}\left(2|3\right)$. Once we have the action
of the charges on a multi-excitation state, we can determine the commutator
of two supercharges, as a check of our results.

In this subsector the charges (\ref{Suparcharges with oscillator})
become\begin{eqnarray}
Q_{a}^{\,\alpha} & = & \sqrt{\frac{m}{3}}\mathrm{Tr}\left\{ \psi^{\alpha}\frac{\partial}{\partial\phi^{a}}-\left(\sqrt{\frac{3}{m}}\right)^{3}\sqrt{2}\varepsilon_{ab}\varepsilon^{\alpha\beta}\left[Z,\phi^{b}\right]\frac{\partial}{\partial\psi^{\beta}}\right\} ,\nonumber \\
S_{\,\alpha}^{a} & = & \sqrt{\frac{m}{3}}\mathrm{Tr}\left\{ -\phi^{a}\frac{\partial}{\partial\psi^{\alpha}}-\left(\sqrt{\frac{3}{m}}\right)^{3}\sqrt{2}\varepsilon^{ab}\varepsilon_{\alpha\beta}\psi^{\beta}\left[\frac{\partial}{\partial Z},\frac{\partial}{\partial\phi^{b}}\right]\right\} ,\end{eqnarray}
 where we chose a coherent state basis, such that\begin{eqnarray*}
c^{\dagger\alpha} & \rightarrow & \psi^{\alpha}\;;\quad c_{\alpha}\rightarrow\frac{\partial}{\partial\psi^{\alpha}};\\
b^{\dagger a} & \rightarrow & \phi^{a}\;;\quad b_{a}\rightarrow\frac{\partial}{\partial\phi^{a}}.\end{eqnarray*}
 For $a=3$, we have the identification $\phi^{3}\equiv Z$. The factor
$\sqrt{\frac{m}{3}}$ will appear as an overall factor in every charge
calculated, and will be dropped, as we know that the quadratic terms
come from the free theory $g_{YM}=0$.

We now proceed to determine the action of the supercharge $Q$ (and
equivalently $S$) on a \emph{single excitation} state $\ket{\chi;J}=\sum_{n}e^{ipn}\ket{Z^{n-1}\chi Z^{J-n+1}}$.
If the excitation is bosonic, $\chi_{\ell}=\phi^{\ell}$, then\[
Q_{a}^{\,\alpha}\ket{\chi;J}=\sum_{n}e^{ipn}\delta_{a}^{\ell}\ket{Z^{n-1}\psi^{\alpha}\left(n\right)Z^{J-n+1};J},\]
 while if the excitation is fermionic, $\chi^{\beta}=\psi^{\beta}$,
we have\begin{eqnarray*}
Q_{a}^{\,\beta}\ket{\chi;J} & = & -\left(\sqrt{\frac{3}{m}}\right)^{3}\sqrt{2}\varepsilon_{ab}\varepsilon^{\alpha\beta}\sqrt{\frac{J+1}{J}}N^{1/2}\ket{Z^{n-1}\left[Z,\phi^{b}\right]Z^{J-n+1};J+1}\\
 & = & -\left(\sqrt{\frac{3}{m}}\right)^{3}\sqrt{2}\varepsilon_{ab}\varepsilon^{\alpha\beta}\sqrt{\frac{J+1}{J}}N^{1/2}\ket{Z^{n}\phi^{b}\left(n+1\right)Z^{J-n+1};J+1}+\\
 &  & \qquad\qquad+\left(\sqrt{\frac{3}{m}}\right)^{3}\sqrt{2}\varepsilon_{ab}\varepsilon^{\alpha\beta}\sqrt{\frac{J+1}{J}}N^{1/2}\ket{Z^{n-1}\phi^{b}\left(n\right)Z^{J-n+2};J+1}\\
 & \approx & -\left(\sqrt{\frac{3}{m}}\right)^{3}\sqrt{2}\varepsilon_{ab}\varepsilon^{\alpha\beta}\sum_{n}e^{ipn}\left(e^{-ip}-1\right)N^{1/2}\ket{Z^{n-1}\phi^{b}\left(n\right)Z^{J-n+2};J+1}\,.\end{eqnarray*}
 It can be seen from the expression above that the insertion of a
$Z$ field before the excitation changes its phase by $e^{-ip}$,
while the insertion after the excitation leaves that phase untouched.
This is a property of the asymptotic state, for which an infinite
number of $Z$ fields exist after the (last) excitation. This was
seen in \cite{Beisert:2005tm} as being equivalent to {}``opening\textquotedbl{}
the trace. In the above expression we also kept only the first order
in $\frac{1}{J}$.

From the results shown above, we can easily determine the generalization
to a \emph{multi-excitation} state. First, rewrite the state as \begin{equation}
\ket{\chi;J}\equiv\ket{\chi_{1}...\chi_{K};J}=\sum_{\left\{ l_{i}\right\} }e^{ip_{1}l_{1}+...+ip_{K}l_{K}}\chi_{1}^{\dagger}\chi_{2}^{\dagger}\cdots\chi_{K}^{\dagger}\ket{0;J}.\end{equation}
 The action of one charge on such state is (zeroth order in $\frac{1}{J}$):

\begin{eqnarray}
Q_{a}^{\,\alpha}\ket{\chi_{1}...\chi_{K};J}\negthickspace\negthickspace & = & \negthickspace\negthickspace\sum_{k=1}^{K}\sum_{\left\{ l_{i}\right\} }e^{ip_{1}l_{1}+...+ip_{K}l_{K}}\left(\prod_{m=1}^{k-1}\left(-1\right)^{F\left(m\right)}\right)\chi_{1}^{\dagger}\chi_{2}^{\dagger}\cdots\left(Q_{a}^{\,\alpha}\chi_{k}^{\dagger}\right)\cdots\chi_{K}^{\dagger}\ket{0;J}\nonumber \\
 & = & \negthickspace\negthickspace\sum_{k=1}^{K}\sum_{\left\{ l_{i}\right\} }e^{ip_{1}l_{1}+...+ip_{K}l_{K}}\left(\prod_{m=1}^{k-1}\left(-1\right)^{F\left(m\right)}\right)\left\{ \delta\left(\chi_{k}^{\dagger},\phi^{b}\right)\delta_{a}^{b}\chi_{1}^{\dagger}\chi_{2}^{\dagger}\cdots\psi^{\alpha}\left(l_{k}\right)\cdots\chi_{K}^{\dagger}\ket{0;J}-\right.\nonumber \\
 &  & \negthickspace\negthickspace-\left.\frac{\sqrt{2N}}{M^{3}}\delta\left(\chi_{k}^{\dagger},\psi^{\beta}\right)\left(\prod_{m=k+1}^{K}e^{-ip_{m}}\right)\left(e^{-ip_{k}}-1\right)\varepsilon_{ab}\varepsilon^{\alpha\beta}\chi_{1}^{\dagger}\chi_{2}^{\dagger}\cdots\phi^{b}\left(l_{k}\right)\cdots\chi_{K}^{\dagger}\ket{0;J+1}\right\} .\nonumber \\
 &  & \qquad\end{eqnarray}
 and similarly for the $S$ charge (noticing that the action of $S$
on a bosonic excitation returns an extra factor of $N$). In here
$\delta\left(\chi_{k}^{\dagger},\phi^{b}\right)$ means that the excitation
$\chi\left(l_{k}\right)$ is bosonic $\phi^{b}$, while in $\delta\left(\chi_{k}^{\dagger},\psi^{\beta}\right)$
the excitation $\chi\left(l{}_{k}\right)$ is fermionic $\psi^{\beta}$.
The factor $\left(-1\right)^{F\left(m\right)}$ is equal to $1$ if
$\chi_{m}$ is bosonic and $-1$ if $\chi_{m}$ is fermionic. Finally
we defined $M=\sqrt{\frac{m}{3}}$. When $\chi_{k}$ is a fermionic
excitation, one gets the expected factor of $\left(e^{-ip_{k}}-1\right)$,
which already showed up in the single excitation case, but one also
gets an extra factor of $\prod_{m=k+1}^{K}e^{-ip_{m}}$. This last
factor can also be explained by the insertion of the $Z$ field. In
fact in the single excitation case we saw that $Z$ changed the momentum
if inserted before the excitation on the chain of fields. But now
the field $Z$ gets inserted before all of the excitations $\chi_{m}$
with $m>k$, hence the change of momenta of all these excitations.

The results of the action of $Q$ and $S$ on a multi-excitation state
will be summarized next using a non local notation (see also \cite{Beisert:2006qh}).

\subsubsection{Twisted vs. non-local notations}

The supercharges $Q$ and $S$ acting on a general state $\ket{\chi;J}$
can be written in a \emph{non-local notation}:\begin{eqnarray}
Q_{a}^{\,\alpha}\ket{\chi;J}\negthickspace & = & \negthickspace\negthickspace\sum_{k=1}^{K}\!\left\{ a_{k}\delta_{a}^{b}\delta\left(\chi_{k}^{\dagger},\phi^{b}\right)\ket{\chi_{1}\cdots\psi^{\alpha}\cdots\chi_{K};J}\!+b_{k}\varepsilon_{ab}\varepsilon^{\alpha\beta}\delta\left(\chi_{k}^{\dagger},\psi^{\beta}\right)\ket{\chi_{1}\cdots\phi^{b}\cdots\chi_{K};J+1}\right\} ,\quad\label{Q action on a state - ak and bk}\\
S_{\,\alpha}^{a}\ket{\chi;J}\negthickspace & = & \negthickspace\negthickspace\sum_{k=1}^{K}\!\left\{ c_{k}\varepsilon^{ab}\varepsilon_{\alpha\beta}\delta\left(\chi_{k}^{\dagger},\phi^{b}\right)\ket{\chi_{1}\cdots\psi^{\beta}\cdots\chi_{K};J-1}\!+d_{k}\delta_{\alpha}^{\beta}\delta\left(\chi_{k}^{\dagger},\psi^{\beta}\right)\ket{\chi_{1}\cdots\phi^{a}\cdots\chi_{K};J}\right\} ,\quad\label{S action on a state - ck and dk}\end{eqnarray}
 where the coefficients are given by\begin{eqnarray}
a_{k}\negthickspace & = & \negthickspace\negthickspace\prod_{m=1}^{k-1}\left(-1\right)^{F\left(m\right)}\,,\nonumber \\
b_{k}\negthickspace & = & \negthickspace\negthickspace\frac{\sqrt{2N}}{M^{3}}\left[\prod_{m=1}^{k-1}\left(-1\right)^{F\left(m\right)}\right]\left(1-e^{-ip_{k}}\right)\left[\prod_{m=k+1}^{K}e^{-ip_{m}}\right]=\frac{\sqrt{2N}}{M^{3}}e^{-iP}\left(e^{ip_{k}}-1\right)\left[\prod_{m=1}^{k-1}\left(-1\right)^{F\left(m\right)}e^{ip_{m}}\right],\nonumber \\
c_{k}\negthickspace & = & \negthickspace\negthickspace\frac{\sqrt{2N}}{M^{3}}\left[\prod_{m=1}^{k-1}\left(-1\right)^{F\left(m\right)}\right]\left(e^{ip_{k}}-1\right)\left[\prod_{m=k+1}^{K}e^{ip_{m}}\right]=\frac{\sqrt{2N}}{M^{3}}e^{iP}\left(1-e^{-ip_{k}}\right)\left[\prod_{m=1}^{k-1}\left(-1\right)^{F\left(m\right)}e^{-ip_{m}}\right],\qquad\label{expressions for ak,bk,ck,dk}\\
d_{k}\negthickspace & = & \negthickspace\negthickspace-\prod_{m=1}^{k-1}\left(-1\right)^{F\left(m\right)}\,.\nonumber \end{eqnarray}

There is one other notation, introduced by Beisert in \cite{Beisert:2006qh},
called the \emph{twisted notation}. In this local notation we have\begin{eqnarray}
Q_{a,k}^{\alpha}\ket{\cdots\phi_{k}^{b}\cdots} & = & a_{k}^{'}\delta_{a}^{b}\ket{\cdots\mathcal{Y}^{+}\psi_{k}^{\alpha}\cdots},\nonumber \\
Q_{a,k}^{\alpha}\ket{\cdots\psi_{k}^{\beta}\cdots} & = & b'_{k}\varepsilon^{\alpha\beta}\varepsilon_{ab}\ket{\cdots\mathcal{Z}^{+}\mathcal{Y}^{-}\phi_{k}^{b}\cdots},\label{Twisted notation: Supercharges}\\
S_{\alpha,k}^{a}\ket{\cdots\phi_{k}^{b}\cdots} & = & c'_{k}\varepsilon^{ab}\varepsilon_{\alpha\beta}\ket{\cdots\mathcal{Z}^{-}\mathcal{Y}^{+}\psi_{k}^{\beta}\cdots},\nonumber \\
S_{\alpha,k}^{a}\ket{\cdots\psi_{k}^{\beta}\cdots} & = & d'_{k}\delta_{\alpha}^{\beta}\ket{\cdots\mathcal{Y}^{-}\phi_{k}^{a}\cdots}.\nonumber \end{eqnarray}
 We notice the presence of the markers $\mathcal{Z}^{\pm},\mathcal{Y^{\pm}}$.
These markers have a simple explanation, up to one loop. The marker
$\mathcal{Y}^{\pm}$ marks the position on the string of fields (the
state) where a fermion field was inserted ($\mathcal{Y}^{+}$) or
removed ($\mathcal{Y}^{-}$). In the twisted notation we are only
given the action of the supercharge on the field $k$ of the string.
But in order for a supercharge to act on such field it will have to
pass by the previous ones. If these are bosonic fields nothing happens,
but if they are fermionic, a minus sign will appear (for each fermionic
fields it passes). Thus, it is important to know where the supercharge
acted, which is done by the marker. The marker is shifted around as
follows:\[
\ket{\cdots\chi_{k}\mathcal{Y}^{\pm}\cdots}=\left(\xi_{k}\right)^{\pm1}\ket{\cdots\mathcal{Y}^{\pm}\chi_{k}\cdots},\]
 where \begin{eqnarray}
\xi_{k} & = & \left(-1\right)^{F\left(k\right)}=\left\{ \begin{array}{l}
1\quad\;\mbox{if }\chi_{k}\,\mbox{bosonic}\\
-1\quad\mbox{if }\chi_{k}\,\mbox{fermionic}\end{array}\right.\,.\label{Definition of squigle of Beisert}\end{eqnarray}

The marker $\mathcal{Z}^{\pm}$ marks a position where an extra $Z$
field was inserted in the string. This changes the length of the vacuum
spin chain, reflecting a change in the momenta of the excitation fields.
But this change in momenta only affects the excitation fields after
the position of the marker. The marker has the property \begin{equation}
\ket{\cdots\chi_{k}\mathcal{Z}^{\pm}\cdots}=\frac{x_{k}^{\pm}}{x_{k}^{\mp}}\ket{\cdots\mathcal{Z}^{\pm}\chi_{k}\cdots},\:\mbox{where}\;\frac{x_{k}^{\pm}}{x_{k}^{\mp}}=e^{\pm ip_{k}},\label{Definition of x+/x- of Beisert}\end{equation}
 with $p_{k}$ being the momenta of the excitation $\chi_{k}$, as
before.

In summary, the \emph{twisted notation} is a local notation, since
it only provides the action of the supercharge on the excitation field
$\chi_{k}$, plus a set of markers that allow us to rewrite it in
a \emph{non-local notation}, as found in (\ref{Q action on a state - ak and bk},
\ref{S action on a state - ck and dk}). We can go from the twisted
notation to the non-local one by removing the markers from the first,
i.e., shifting them so that they will be at the right (or left) of
all the excitation fields.

In the local \emph{twisted notation} we have%
\footnote{The coupling constant $M^{6}=\left(\frac{m}{3}\right)^{3}$ is related
to the Yang-Mills coupling constant $g_{YM}$ in the following way\[
\frac{1}{M^{6}}=\frac{g_{YM}^{2}}{32\pi^{2}}.\]
 This relation comes from matching the prefactor of the reduced SYM
action with the prefactor of the matrix model action. In fact we had
$m=\frac{6}{R}$, where $R$ was the radius of $S^{3}$. Taking the
radius small corresponds to $m\gg1$ and consequently $g_{YM}\ll1$. %
} \begin{eqnarray*}
a'_{k} & = & -d'_{k}=1\,,\quad b'_{k}=\frac{\sqrt{2N}}{M^{3}}\left(1-e^{-ip_{k}}\right),\quad c'_{k}=-\frac{\sqrt{2N}}{M^{3}}\left(1-e^{ip_{k}}\right)\,.\end{eqnarray*}

\subsubsection{Comparison with Beisert at 1-loop}

One can find the all-loop version of these coefficients in \cite{Beisert:2006qh},
for both the non-local and the twisted notation. In fact we can expand
the (non-local) coefficients given in that reference to order $\mathcal{O}\left(g\right)$,
and compare them to our results. These coefficients are:\begin{eqnarray*}
a_{k} & = & \gamma_{k}\prod_{j=1}^{k-1}\left(-1\right)^{F\left(j\right)},\\
b_{k} & = & g\frac{\alpha}{\gamma_{k}}\left(1-e^{ip_{k}}\right)\prod_{j=1}^{k-1}\left(e^{ip_{k}}\left(-1\right)^{F\left(j\right)}\right),\\
c_{k} & = & i\frac{\gamma_{k}}{\alpha x_{k}^{+}}\prod_{j=1}^{k-1}\left(e^{-ip_{k}}\left(-1\right)^{F\left(j\right)}\right),\\
d_{k} & = & g\frac{x_{k}^{+}}{i\gamma_{k}}\left(1-e^{-ip_{k}}\right)\prod_{j=1}^{k-1}\left(-1\right)^{F\left(j\right)}.\end{eqnarray*}
 We used the identifications (\ref{Definition of squigle of Beisert})
and (\ref{Definition of x+/x- of Beisert}) into the transcribed coefficients,
and also made a rescaling of the parameter $\gamma_{k}\rightarrow\sqrt{g}\gamma_{k}$.
The expansion in $g$ is hidden in the dependence of $x^{+},x^{-}$
on the coupling constant:\begin{equation}
x^{+}+\frac{1}{x^{+}}-x^{-}-\frac{1}{x^{-}}=\frac{i}{g}.\end{equation}
 This last equation, together with (\ref{Definition of x+/x- of Beisert}),
allows us to solve for $x^{+}\left(g\right)$:\[
x^{+}=i\frac{1+\sqrt{1+16g^{2}\sin^{2}\left(p/2\right)}}{2g\left(1-e^{-ip}\right)}.\]
 Then by expanding this expression up to order $\mathcal{O}\left(g\right)$,
we obtain exact agreement with (\ref{expressions for ak,bk,ck,dk}),
as long as we identify $\gamma_{k}=\left(-1\right)^{F\left(k\right)}$
and $\alpha=e^{-iP}$. Note that the relation between the normalized
't Hooft coupling $g$ and the Yang-Mills coupling constant $g_{YM}$
is $g=\frac{g_{YM}}{4\pi}\sqrt{N_{c}}$, from the gauge group $SU\left(N_{c}\right)$.

The other charges that we are interested in determining are the hamiltonian
$H$ and the central charges of the extended algebra $P,K$. These
charges arise from commutation relations between the supercharges,
which will be determined next.

\subsection{Commutation Relations}

At this moment we have calculated only the supercharges of the full
extended algebra $\mathfrak{su}\left(2|2\right)$, up to $\mathcal{O}\left(g\right)$.
We are interested in having the complete set of charges at this order,
which comprises also the rotations generators $L,R$, the dilatation
operator $H$, and also the central charges of the extended algebra
$P,K$ (bosonic generators of momentum and boosts, which have zero
eigenvalues when applied to physical states). All of these generators
can be obtained to $\mathcal{O}\left(g\right)$ from the commutation
relations of the supercharges.

The central charges of the extended algebra receive no loop corrections
, and as such, can be obtained exactly by the anti-commutation relations
$\left\{ Q,Q\right\} \sim P$ and $\left\{ S,S\right\} \sim K$, by
knowing the zeroth order of the supercharges. The other generators
will be obtained from the last anti-commutator $\left\{ Q,S\right\} \propto R+L+H$,
but while the zeroth order supercharges will be enough to determine
rotation generators $L$ and $R$, the central charge $H$ will only
be know correctly up to $\mathcal{O}\left(g\right)$, as we'll see
below.

In the anti-commutator of any two supercharges the only terms that
will not vanish are the ones where the two supercharges are applied
to the same excitation. The anti-commutator of two $Q$ charges is:

\begin{eqnarray*}
\left\{ Q_{b}^{\,\beta},Q_{a}^{\,\alpha}\right\} \ket{\chi_{1}...\chi_{K};J} & = & \sum_{k=1}^{K}\ket{\chi_{1}\cdots\left(\left\{ Q_{b}^{\,\beta},Q_{a}^{\,\alpha}\right\} \chi_{k}\right)\cdots\chi_{K};J}\\
 & = & \frac{\sqrt{2N}}{M^{3}}\sum_{k=1}^{K}\left[\left(1-e^{-ip_{k}}\right)\prod_{l=k+1}^{K}e^{-ip_{l}}\right]\ket{\chi_{1}...\chi_{K};J+1}\\
 & = & \frac{\sqrt{2N}}{M^{3}}\left(1-e^{-i\sum_{k=1}^{K}p_{k}}\right)\ket{\chi_{1}...\chi_{K};J+1}.\end{eqnarray*}
 This is just the action of the central charge $\left\{ Q,Q\right\} \propto P$
of the extended algebra on a multi-excitation state. The action of
the other central charge of the extended algebra is obtained from
$\left\{ S,S\right\} \propto K$:\[
\left\{ S_{\,\beta}^{b},S_{\,\alpha}^{a}\right\} \ket{\chi_{1}...\chi_{K};J}=\frac{\sqrt{2N}}{M^{3}}\left(1-e^{i\sum_{k=1}^{K}p_{k}}\right)\ket{\chi_{1}...\chi_{K};J-1}.\]

We know from \cite{Beisert:2006qh} that there is an outer automorphism
relating $H$ and the central charges of the extended algebra $P,K$,
which corresponds to an $\mathfrak{sl}\left(2\right)$ algebra. Closure
of this algebra on the original commutation relations of the supercharges
requires that\begin{equation}
H^{2}-PK=\frac{1}{4}.\end{equation}
 This relation should only hold when we consider the all loop $H$,
and not only when we consider the first two orders. Using non-local
notation, we find that the product $PK$ is given by\[
PK=-\frac{2N}{M^{6}}\left(e^{-i\sum_{k=1}^{K}p_{k}}-1\right)\left(e^{+i\sum_{k=1}^{K}p_{k}}-1\right)=\frac{8N}{M^{6}}\sin^{2}\left(\sum_{k=1}^{K}\frac{p_{k}}{2}\right)=\frac{8N}{M^{6}}\sin^{2}\left(\frac{p}{2}\right),\]
 and so $H^{2}=\frac{1}{4}+PK=\frac{1}{4}+\frac{8N}{M^{6}}\sin^{2}\left(\frac{p}{2}\right)$,
which implies\[
H=\pm\frac{1}{2}\sqrt{1+\frac{32N}{M^{6}}\sin^{2}\left(\frac{p}{2}\right)}=\pm\frac{1}{2}\sqrt{1+\frac{g_{YM}^{2}N}{\pi^{2}}\sin^{2}\left(\frac{p}{2}\right)}\,.\]
 This is the result expected at one loop. The identification of the
matrix model mass parameter with the Yang-Mills coupling
holds at one loop but some mismatches were seen to appear at higher
loop calculations, implying some kind of BMN scaling breakdown, and
a substitution of the factor $\frac{32N}{M^{6}}$ for a function $f\left(\frac{N}{M^{6}}\right)$
\cite{Fischbacher:2004iu}.

We now calculate the anti-commutator of $Q$ and $S$, which will
be proportional to $L_{\beta}^{\,\alpha},\, R_{\, b}^{a}$ and
the hamiltonian $H$:\begin{eqnarray*}
\left\{ Q_{b}^{\,\beta},S_{\,\alpha}^{a}\right\} \ket{\chi_{1}...\chi_{K};J} & = & \sum_{k=1}^{K}\left\{ c_{k}b_{k}\varepsilon^{aa'}\varepsilon_{bb'}\delta_{\alpha}^{\beta}\delta\left(\chi_{k}^{\dagger},\phi^{a'}\right)\ket{\chi_{1}\cdots\phi^{b'}\cdots\chi_{K};J}+\right.\\
 &  & \qquad\quad+c_{k}b_{k}\varepsilon^{\beta\beta'}\varepsilon_{\alpha\alpha'}\delta_{b}^{a}\delta\left(\chi_{k}^{\dagger},\psi^{\beta'}\right)\ket{\chi_{1}\cdots\psi^{\alpha'}\cdots\chi_{K};J}+\\
 &  & \qquad\quad+a_{k}d_{k}\delta_{\alpha}^{\beta}\delta\left(\chi_{k}^{\dagger},\phi^{b}\right)\ket{\chi_{1}\cdots\phi^{a}\cdots\chi_{K};J}+\\
 &  & \qquad\quad\left.+a_{k}d_{k}\delta_{b}^{a}\delta\left(\chi_{k}^{\dagger},\psi^{\alpha}\right)\ket{\chi_{1}\cdots\psi^{\beta}\cdots\chi_{K};J}\right\} \,.\end{eqnarray*}
 From equations (\ref{expressions for ak,bk,ck,dk}) we have that:\[
a_{k}d_{k}=-1\qquad;\qquad b_{k}c_{k}=\frac{4N}{M^{6}}\left(1-e^{-ip_{k}}\right)\left(e^{ip_{k}}-1\right)=-\frac{16N}{M^{6}}\sin^{2}\left(\frac{p_{k}}{2}\right).\]
 Also, we know from the algebra \eqref{Commutation relations of symmetry generators}
that \begin{eqnarray*}
\mathcal{L}_{\alpha}^{\beta}\ket{\psi^{\gamma}} & = & \delta_{\alpha}^{\gamma}\ket{\psi^{\beta}}-\frac{1}{2}\delta_{\alpha}^{\beta}\ket{\psi^{\gamma}},\\
R_{b}^{a}\ket{\phi^{c}} & = & \delta_{b}^{c}\ket{\phi^{a}}-\frac{1}{2}\delta_{b}^{a}\ket{\phi^{c}}.\end{eqnarray*}
 For multi-particle states this generalizes to \[
\mathcal{L}_{\alpha}^{\beta}\ket{\chi_{1}...\chi_{K};J}=\sum_{k=1}^{K}\chi_{1}^{\dagger}\cdots\mathcal{L_{\alpha}^{\beta}}\left(\chi_{k}^{\dagger}\right)\cdots\chi_{K}^{\dagger}\ket{0;J},\]
 with a similar result for the charge $R_{b}^{a}$.%
\footnote{The charges $\mathcal{L}$ and $R$ are the generators of the algebra that
correspond to rotations of the $\psi^{\gamma}$ $\mathfrak{su}\left(2\right)$
algebra and of the $\phi^{a}$ $\mathfrak{su}\left(2\right)$ algebras,
respectively. As such, $\mathcal{L}_{\alpha}^{\beta}\ket{\phi^{c}}=0$,
and $R_{b}^{a}\ket{\psi^{\gamma}}=0$.%
}

One can now easily see that\begin{eqnarray}
\left\{ Q_{b}^{\,\beta},S_{\,\alpha}^{a}\right\} \ket{\chi_{1}...\chi_{K};J} & \negthickspace=\negthickspace & \delta_{\alpha}^{\beta}R_{b}^{a}\ket{\chi_{1}...\chi_{K};J}+\delta_{b}^{a}\mathcal{L}_{\alpha}^{\beta}\ket{\chi_{1}...\chi_{K};J}+\delta_{\alpha}^{\beta}\delta_{b}^{a}\sum_{k=1}^{K}\left(\frac{1}{2}a_{k}d_{k}+b_{k}c_{k}\right)\ket{\chi_{1}...\chi_{K};J}\nonumber \\
 &  & -\delta_{\alpha}^{\beta}\sum_{k=1}^{K}b_{k}c_{k}\delta\left(\chi_{k}^{\dagger},\phi^{b}\right)\ket{\chi_{1}\cdots\phi^{a}\cdots\chi_{K};J}\nonumber \\
 &  & -\delta_{b}^{a}\sum_{k=1}^{K}b_{k}c_{k}\delta\left(\chi_{k}^{\dagger},\psi^{\alpha}\right)\ket{\chi_{1}\cdots\psi^{\beta}\cdots\chi_{K};J}.\label{Commutator of {Q,S}}\end{eqnarray}
 If we compare with the expected results from commutation relations
given in \eqref{Commutation relations of symmetry generators}, the
last two terms seem to be extra. But in fact this is the exact result!
We (anti-)commuted only the order $g^{0}$ and order $g^{1}$ of the
supercharges. That is, we calculated the nonzero anti-commutators
$\left\{ Q_{0},S_{0}\right\} \propto R+L+H_{0}$ and $\left\{ Q_{1},S_{1}\right\} $.
This last anti-commutator contributes to order $g^{2}$ of the hamiltonian,
$H_{2}$, but there will be another contribution to $H_{2}$: the
two-loop terms of the supercharges, $Q_{2}$ and $S_{2}$, will have
nonzero commutation relations with $S_{0}$ and $Q_{0}$, respectively,
and contribute to $\mathcal{O}\left(g^{2}\right)$. So $H_{2}$(the
energy central charge of order $g^{2}$) will be fully determined
by:\begin{equation}
H_{2}\propto\left\{ S_{1},Q_{1}\right\} +\left\{ S_{2},Q_{0}\right\} +\left\{ S_{0},Q_{2}\right\} .\label{two-loop Hamiltonian}\end{equation}
 Only considering all the above anti-commutators we will get the correct
result for the $H_{2}$. For calculations see Appendix \ref{sec:Commuting-the-charges-2-loops},
and also \cite{Beisert:2003ys}.

\subsection{Supercharges as operators in momentum space\label{sub:Oscillator_generators_SYM}}

We now present a description of the supercharges in terms of operators
in momentum space. Consider as before an infinite chain of fields
$Z$. The vacuum state, written before as $\ket{0;J}=\mbox{Tr}\left(Z^{J}\right)\ket{0},$
can be rewritten, in the {}``hamiltonian formalism'' introduced
in \cite{Berenstein:2002jq} as $\ket{0;J}=\left(b_{z}^{\dagger}\right)^{J}\ket{0}$,
where $b_{z}^{\dagger}$ creates an extra $Z$ field in the string.%
\footnote{The subscript $z$ is used in this section, to distinguish the creation
operator $b_{z}^{\dagger}$ for the boson $Z$ from the creation operator
$b^{a\dagger}$ for the two bosonic impurities.%
} Then we can write a state with $K$ impurities as: \[
\ket{\Psi}=\sum_{n_{1},...,n_{K}}e^{ip_{j}n_{j}}b^{\dagger}\left(n_{1}\right)\cdots b^{\dagger}\left(n_{K}\right)\ket{0;J}=b^{\dagger}\left(p_{1}\right)\cdots b^{\dagger}\left(p_{K}\right)\ket{0;J}.\]
 We are imposing dilute gas approximation, in which we consider $n_{1}\ll n_{2}\ll\cdots\ll n_{K}$.
We will now assume $p_{1}<p_{2}<\cdots<p_{K}$.

In the last expression for $\ket{\Psi}$ we used the creation operators
$b^{\dagger}\left(n\right)=\left(b_{z}^{\dagger}\right)^{n}b^{\dagger}\left(b_{z}\right)^{n}$,
which create a boson $b$ at position $n$ in the string of $Z's$.
One can also introduce $c^{\dagger}\left(n\right)=\left(b_{z}^{\dagger}\right)^{n}c^{\dagger}\left(b_{z}\right)^{n}$
as a creation operator for a fermion at position $n$. The action
of the hamiltonian in this framework can be found in \cite{Berenstein:2002jq},
and a further comparison with lattice strings can be found in \cite{deMelloKoch:2003pv}.

To write the action of the supercharges in terms of these operators,
we also need to introduce a partial momentum operator\[
\hat{\mathcal{P}}\left(p\right)=\int_{0}^{p}dp'\, p'\left[b^{\dagger}\left(p'\right)b\left(p'\right)+c^{\dagger}\left(p'\right)c\left(p'\right)\right],\]
 or the discrete momentum version\[
\hat{\mathcal{P}}\left(p\right)=\sum_{k=0}^{p-1}k\left[b^{\dagger}\left(k\right)b\left(k\right)+c^{\dagger}\left(k\right)c\left(k\right)\right].\]
 The total momentum operator is just $\hat{P}=\hat{\mathcal{P}}\left(p_{max}\right)$,
where $p_{max}$ is either $\infty$ in the continuum case, or finite
(but large) in the lattice. Also, define an operator $\hat{\Theta}$
conjugate to the {}``R-charge operator\textquotedbl{} $\hat{\mathcal{J}\,}$.
In the spin-chain formalism, $\hat{\mathcal{J}\,}$ effectively measures
the length of the chain of $Z$ fields, and $\hat{\Theta}$ changes
that length:

\[
\hat{\mathcal{J}\,}e^{\pm i\hat{\Theta}}\ket{0,J}=\left(J\pm1\right)e^{\pm i\hat{\Theta}}\ket{0,J}.\]

We can now proceed to the action of the supercharges. In momentum
space, they become:\begin{eqnarray}
Q_{b}^{\beta} & = & -\frac{\sqrt{2N}}{M^{3}}\varepsilon_{bb'}\varepsilon^{\beta\beta'}e^{i\hat{\Theta}}e^{-i\hat{P}}\sum_{p}b^{b'\dagger}\left(p\right)\left(e^{ip}-1\right)e^{i\hat{\mathcal{P}}\left(p\right)}c_{\beta'}\left(p\right)+\sum_{p}c^{\beta\dagger}\left(p\right)b_{b}\left(p\right);\nonumber \\
S_{\alpha}^{a} & = & \frac{\sqrt{2N}}{M^{3}}\varepsilon_{\alpha\alpha'}\varepsilon^{aa'}e^{-i\hat{\Theta}}e^{i\hat{P}}\sum_{p}c^{\alpha'\dagger}\left(p\right)\left(1-e^{-ip}\right)e^{-i\hat{\mathcal{P}}\left(p\right)}b_{a'}\left(p\right)-\sum_{p}b^{a\dagger}\left(p\right)c_{\alpha}\left(p\right).\label{Supercharges in terms of osc}\end{eqnarray}
 It is not hard to check that these definition give us the results
obtained in the previous section. In the above expression the sum
over momenta has increments of $\frac{2\pi}{J}$.%
\footnote{ The operator $e^{\pm i\hat{\Theta}}$ does not commute with the sum
over the momenta, as it changes the increments in the sum. But in
the limit $J$ very large, this change will be negligible.%
}

Commuting two central charges $Q$ will give us the central charge
$\mathcal{P}$:\begin{equation}
\left\{ Q,Q\right\} =e^{i\hat{\Theta}}e^{-i\hat{P}}\sum_{p}b^{\dagger}\left(p\right)\left(e^{ip}-1\right)e^{i\hat{\mathcal{P}}\left(p\right)}b\left(p\right)+e^{i\hat{\Theta}}e^{-i\hat{P}}\sum_{p}c^{\dagger}\left(p\right)\left(e^{ip}-1\right)e^{i\hat{\mathcal{P}}\left(p\right)}c\left(p\right)=\mathcal{P}.\end{equation}
 One can show that the central charge takes the much more common form:%
\footnote{It can be proven by using the property (valid for any power $n$,
proven by induction, and for $\chi$ fermionic or bosonic)\[
\hat{\mathcal{P}}^{n}\left(p\right)\chi^{\dagger}\left(p'\right)=\chi^{\dagger}\left(p'\right)\left[\theta\left(p-p'\right)p'+\hat{\mathcal{P}}\left(p\right)\right]^{n}.\]
}\begin{equation}
\mathcal{P}=e^{i\hat{\Theta}}e^{-i\hat{P}}\left(e^{i\hat{P}}-1\right)=e^{i\hat{\Theta}}\left(1-e^{-i\hat{P}}\right).\end{equation}
 To summarize, we found expressions for the supercharges as operators
in momentum space, as well as for their commutation relations, in
the large $J$ limit. These expressions once applied to states with
$K$ impurities will result in the expressions obtained in the previous
section.

\section{SUSY generators in $AdS_{5}\times S^{5}$\label{sec:SUSY-generators-in AdSxS}}

This section will be devoted to determining the action of the supercharges
from the string action in $AdS_{5}\times S^{5}$ on a lattice string,
followed by a comparison of its structure to supercharges actions obtained
in the previous section \ref{sub:Oscillator_generators_SYM}.

We now turn to the action of the supercharges from the $AdS_{5}\times S^{5}$
perspective. We start from the results of \cite{Arutyunov:2006ak,Arutyunov:2006yd}.
In Appendix \ref{sec:The-Q-charges-string} we find a summary of those
results, and their restriction to the $\mathfrak{su}\left(2|2\right)$
subsector. The fermionic supercharges $Q$ and $S$ are given by:\begin{eqnarray*}
S_{\, a}^{\alpha} & = & -\frac{1}{2}\int d\sigma e^{-\frac{i}{2}x_{-}}\left(i\theta^{\alpha}\left(2P^{Y}+iY\right)_{a}-\epsilon^{\alpha\beta}\epsilon_{ab}\theta_{\beta}^{\dagger}Y'^{b}\right),\\
Q_{\alpha}^{\, a} & = & \frac{1}{2}\int d\sigma e^{\frac{i}{2}x_{-}}\left(i\theta_{\alpha}^{\dagger}\left(2P^{Y}-iY\right)^{a}+\epsilon_{\alpha\beta}\epsilon^{ab}\theta^{\beta}Y'_{b}\right).\end{eqnarray*}
 Before continuing, let us notice that the coordinate $x_{-}\left(\sigma\right)$
obeys:\[
x_{-}\left(\sigma\right)=\int_{-r}^{\sigma}d\sigma'x_{-}^{'}\left(\sigma'\right)+x_{-}\left(-r\right)=\int_{-r}^{\sigma}d\sigma'\pi_{ws}\left(\sigma'\right)+x_{-}^{0},\]
 where $x'_{-}=\pi_{ws}\left(\sigma\right)$ is the worldsheet momentum
density. The total worldsheet momentum is given by $p_{ws}=\int_{-r}^{r}d\sigma\pi_{ws}\left(\sigma\right)$.

We now want to perform a mode expansion. To do so we will follow the
notation of \cite{Frolov:2006cc}. For the bosonic fields we have:\begin{eqnarray}
Y_{a} & = & \frac{1}{\sqrt{\omega}}\left(A_{a}+B_{a}^{\dagger}\right)\quad;\quad P^{a}=\frac{\sqrt{\omega}}{4i}\left(A^{\dagger a}-B^{a}\right);\nonumber \\
Y^{a}=\overline{Y}_{a} & = & \frac{1}{\sqrt{\omega}}\left(A^{\dagger a}+B^{a}\right)\quad;\quad P_{a}=\overline{P}^{a}=i\frac{\sqrt{\omega}}{4}\left(A_{a}-B_{a}^{\dagger}\right),\end{eqnarray}
 where $\omega=\sqrt{1+\frac{1}{2}\tilde{\lambda}\partial_{\sigma}^{2}}$,
and $\tilde{\lambda}$ is the effective coupling constant (light-cone
gauge) in the pp-wave limit $\tilde{\lambda}\equiv\frac{4\lambda}{P_{+}^{2}}$,
kept finite when $P_{+},\lambda\rightarrow\infty$. For the fermionic
fields we have:\begin{equation}
\theta^{\alpha}=\sqrt{\frac{1}{2}\left(1+\frac{1}{\omega}\right)}c^{\alpha}\quad;\qquad\theta_{\alpha}^{\dagger}=\sqrt{\frac{1}{2}\left(1+\frac{1}{\omega}\right)}c_{\alpha}^{\dagger}.\end{equation}
 With these expansions, we get the following results:\begin{eqnarray*}
i\theta^{\alpha}\left(2P^{Y}+iY\right)_{a} & = & -\sqrt{\frac{\omega+1}{2\omega}}c^{\alpha}\left\{ \frac{\sqrt{\omega}}{2}\left(A_{a}-B_{a}^{\dagger}\right)+\frac{1}{\sqrt{\omega}}\left(A_{a}+B_{a}^{\dagger}\right)\right\} ,\\
\theta_{\beta}^{\dagger}Y'^{b} & = & \sqrt{\frac{\omega+1}{2\omega}}c_{\alpha}^{\dagger}\frac{\sqrt{\tilde{\lambda}}\partial_{\sigma}}{\sqrt{2\omega}}\left(A^{\dagger b}+B^{b}\right).\end{eqnarray*}

We will be keeping $Y\approx B^{\dagger}$, dropping the oscillators
$A,A^{\dagger}$. Then up to order $\mathcal{O}\left(\sqrt{\tilde{\lambda}}\right)$,\begin{equation}
Q_{\,\alpha}^{a}=\frac{1}{4}\int d\sigma e^{\frac{i}{2}x_{-}}\left(c_{\alpha}^{\dagger}B^{a}+\sqrt{2}\epsilon_{\alpha\beta}\epsilon^{ab}c^{\beta}\sqrt{\tilde{\lambda}}\partial_{\sigma}B_{b}^{\dagger}\right).\end{equation}
 The same can be done for the supercharge $S$, which then becomes:\begin{equation}
S_{\, a}^{\alpha}=\frac{1}{4}\int d\sigma e^{-\frac{i}{2}x_{-}}\left(c^{\alpha}B_{a}^{\dagger}+\sqrt{2}\epsilon^{\alpha\beta}\epsilon_{ab}c_{\beta}^{\dagger}\sqrt{\tilde{\lambda}}\partial_{\sigma}B^{b}\right).\end{equation}

For a comparison with the Super Yang-Mills supercharges found in \prettyref{sub:Oscillator_generators_SYM},
we need to discretize the above results. To do so recall that $r=P_{+}/2$,
and $\int_{-r}^{r}d\sigma=P_{+}$. Then the lattice version of $Q$
is:\begin{eqnarray}
Q_{\,\alpha}^{a} & = & \frac{1}{4}\sum_{\ell=1}^{P_{+}}e^{ix_{-}^{0}/2}\left(\prod_{k=0}^{\ell}e^{\frac{i}{2}\pi\left(k\right)}\right)\left\{ c_{\alpha}^{\dagger}\left(\ell\right)B^{a}\left(\ell\right)+\sqrt{2}\epsilon_{\alpha\beta}\epsilon^{ab}c^{\beta}\left(\ell\right)\sqrt{\tilde{\lambda}}\left(B_{b}^{\dagger}\left(\ell\right)-B_{b}^{\dagger}\left(\ell-1\right)\right)\right\} \nonumber \\
 & = & \frac{1}{4}\sum_{\ell=1}^{P_{+}}e^{\frac{i}{2}x_{-}^{0}}e^{\frac{i}{2}p\left(\ell\right)}\left\{ c_{\alpha}^{\dagger}\left(\ell\right)B^{a}\left(\ell\right)+\sqrt{2}\epsilon_{\alpha\beta}\epsilon^{ab}\sqrt{\tilde{\lambda}}\left(B_{b}^{\dagger}\left(\ell\right)-B_{b}^{\dagger}\left(\ell-1\right)\right)c^{\beta}\left(\ell\right)\right\} ,\end{eqnarray}
 where $p\left(\ell\right)=\sum_{k=1}^{\ell}\pi\left(k\right)$.

To continue, we need to write what $p\left(\ell\right)$ does to an
excitation: \[
e^{\frac{i}{2}p\left(\ell\right)}\chi\left(\ell_{k}\right)e^{-\frac{i}{2}p\left(\ell\right)}=\left\{ \begin{array}{l}
\chi\left(\ell_{k}\right)\quad\qquad\ell_{k}<\ell\\
\chi\left(\ell_{k}+1\right)\quad\;\ell_{k}>\ell\end{array}\right.\,.\]
 By performing the following change of variables, $c_{\alpha}^{\dagger}\left(\ell\right)\rightarrow e^{-\frac{i}{2}x_{-}^{0}}e^{-\frac{i}{2}p\left(\ell\right)}c_{\alpha}^{\dagger}\left(\ell\right)$,
the charge becomes:\begin{eqnarray}
Q_{\,\alpha}^{a} & = & \frac{1}{4}\sum_{\ell=1}^{P_{+}}\left\{ c_{\alpha}^{\dagger}\left(\ell\right)B^{a}\left(\ell\right)+\sqrt{2}\epsilon_{\alpha\beta}\epsilon^{ab}\sqrt{\tilde{\lambda}}e^{\frac{i}{2}x_{-}^{0}}e^{\frac{i}{2}p\left(\ell\right)}c^{\beta}\left(\ell\right)e^{\frac{i}{2}p\left(\ell\right)}e^{\frac{i}{2}x_{-}^{0}}\left(B_{b}^{\dagger}\left(\ell\right)-B_{b}^{\dagger}\left(\ell-1\right)\right)\right\} \nonumber \\
 & = & \frac{1}{4}\sum_{\ell=1}^{P_{+}}\left\{ c_{\alpha}^{\dagger}\left(\ell\right)B^{a}\left(\ell\right)+\sqrt{2}\epsilon_{\alpha\beta}\epsilon^{ab}\sqrt{\tilde{\lambda}}e^{ix_{-}^{0}}\left(B_{b}^{\dagger}\left(\ell\right)-B_{b}^{\dagger}\left(\ell-1\right)\right)e^{ip\left(\ell\right)}c^{\beta}\left(\ell\right)\right\} .\end{eqnarray}
 The other supercharge $S_{\, a}^{\alpha}$ can also be determined
to be:\begin{eqnarray}
S_{\, a}^{\alpha}\negthickspace & = & \negthickspace\negthickspace\frac{1}{4}\sum_{\ell=1}^{P_{+}}e^{-\frac{i}{2}x_{-}^{0}}e^{-\frac{i}{2}p\left(\ell\right)}\left(c^{\alpha}\left(\ell\right)e^{\frac{i}{2}p\left(\ell\right)}e^{\frac{i}{2}x_{-}^{0}}B_{a}^{\dagger}\left(\ell\right)+\sqrt{2}\epsilon^{\alpha\beta}\epsilon_{ab}e^{-\frac{i}{2}x_{-}^{0}}e^{-\frac{i}{2}p\left(\ell\right)}c_{\beta}^{\dagger}\left(\ell\right)\sqrt{\tilde{\lambda}}\left(B^{b}\left(\ell\right)-B^{b}\left(\ell-1\right)\right)\right)\nonumber \\
 & = & \negthickspace\negthickspace\frac{1}{4}\sum_{\ell=1}^{P_{+}}\left(B_{a}^{\dagger}\left(\ell\right)c^{\alpha}\left(\ell\right)+\sqrt{2}\epsilon^{\alpha\beta}\epsilon_{ab}\sqrt{\tilde{\lambda}}e^{-ix_{-}^{0}}e^{-ip\left(\ell\right)}c_{\beta}^{\dagger}\left(\ell\right)\left(B^{b}\left(\ell\right)-B^{b}\left(\ell-1\right)\right)\right)\,.\end{eqnarray}
 If we wrote these charges in momentum space, we would obtain the
exact structure for the supercharges (\ref{Supercharges in terms of osc}),
as long as we make the correspondence that the conjugate pair $\left(x_{-}^{0},P_{+}\right)\leftrightarrow\left(\hat{\Theta},\hat{\mathcal{J\,}}\right)$.
In the above expressions, $x_{-}^{0}$ plays the part of the length
changing operator, as it is the conjugate variable to $P_{+}$, the
total light-cone momentum, which is in its turn related to the width
of the worldsheet cylinder. For closed strings the total worldsheet
momentum $p_{ws}$ has to vanish (on-shell) - level-matching condition.
If we relax this condition (off-shell) and take $P_{+}\rightarrow\infty$,
then we obtain the centrally extended algebra with extra central charges
$C,C^{*}$ added to the hamiltonian $H$ (the same as the generators
of translations $P$ and boosts $K$).

One other way of checking the results is by writing the supercharges
in first quantized framework. Choosing again a state such that: \[
\ket{\chi_{1}\cdots\chi_{K};P_{+}}=\sum_{\left\{ m_{i}\right\} =0}^{P^{+}}e^{ip_{1}m_{1}+\cdots+ip_{K}m_{K}}\chi_{1}\left(m_{1}\right)\cdots\chi_{K}\left(m_{K}\right)\ket{0;P_{+}},\]
 where $\chi_{i}\left(m_{i}\right)=b_{z}^{m_{i}}\chi_{i}b_{z}^{-m_{i}}$,
with $b_{z}$ being the oscillators equivalent to the field $Z$.
Then\begin{eqnarray}
Q_{\,\alpha}^{a}\ket{\chi_{1}\cdots\chi_{K};P_{+}}\negthickspace\negthickspace & = & \negthickspace\negthickspace\frac{1}{4}\sum_{k=1}^{K}\left(\prod_{m=1}^{k-1}\left(-1\right)^{F\left(m\right)}\right)\left\{ \delta(\chi_{k},B_{b}^{\dagger})\delta_{b}^{a}\ket{\chi_{1}\cdots c_{\alpha}^{\dagger}\left(k\right)\cdots\chi_{K};P_{+}}+\right.\nonumber \\
 &  & \negthickspace\negthickspace\negthickspace\negthickspace+\left.\sqrt{2\,\tilde{\lambda}}\,\delta(\chi_{k},c_{\beta}^{\dagger})\epsilon^{ab}\epsilon_{\alpha\beta}\left(\prod_{l=k+1}^{K}e^{ip_{l}}-\prod_{l=k}^{K}e^{ip_{l}}\right)\ket{\chi_{1}\cdots B_{b}^{\dagger}\left(k\right)\cdots\chi_{K};P_{+}+1}\right\} \negmedspace.\quad\end{eqnarray}
 Doing the same calculation for the $S$ generator, one gets:\begin{eqnarray}
S_{\, a}^{\alpha}\ket{\chi_{1}\cdots\chi_{K};P_{+}}\negthickspace\negthickspace & = & \negthickspace\negthickspace\frac{1}{4}\sum_{k=1}^{K}\left(\prod_{m=1}^{k-1}\left(-1\right)^{F\left(m\right)}\right)\left\{ \delta(\chi_{k},c_{\beta}^{\dagger})\delta_{\beta}^{\alpha}\chi_{1}\left(m_{1}\right)\cdots\left(B_{a}^{\dagger}\left(k\right)\right)\cdots\chi_{K}\left(m_{K}\right)\ket{0;P_{+}}+\right.\nonumber \\
 &  & \negthickspace\negthickspace\negthickspace\negthickspace\negthickspace+\left.\sqrt{2\,\tilde{\lambda}}\,\delta(\chi_{k},B_{b}^{\dagger})\epsilon^{\alpha\beta}\epsilon_{ab}\left(\prod_{l=k+1}^{K}e^{-ip_{l}}-\prod_{l=k}^{K}e^{-ip_{l}}\right)\ket{\chi_{1}\cdots c_{\beta}^{\dagger}\left(k\right)\cdots\chi_{K};P_{+}-1}\right\} \negmedspace.\qquad\end{eqnarray}

From this we can again see that the actions of the supercharges $Q$
and $S$, have a similar structure at one-loop, on both sides of the
correspondence. But while the results presented in this section are
perturbative in $\tilde{\lambda}$ (BMN limit), the results presented
in the previous section are perturbative in the 't Hooft coupling
$\lambda$, so one cannot perform a direct comparison.

\section{Conclusions and Acknowledgments\label{sec:Conclusions}}

In this paper we studied in detail the $Q,\, S$ generators of the
extended algebra $\mathfrak{su}\left(2|2\right)$in the plane-wave
matrix theory formalism. By using a coherent basis we determined the
supercharges in the non-local notation of Beisert \cite{Beisert:2006qh}
(as well as in the local twisted notation), and determined some of
the coefficients in this notation up to order $\mathcal{O}\left(g_{YM}\right)$.

We also determined the anti-commutation relations of these supercharges,
and obtained the expected results for the central charges $P,K$ and
$H$. We saw that we needed to know the hamiltonian up to two-loops
in order to have a closed (anti-)commutation relation between $Q$
and $S$.

We finally wrote a first quantized formulation of the supercharges
obtained directly from the sigma model action for the string. Having
the supercharges written in that way allowed us to compare their structure
with the what we had previously calculated from gauge side.

The evidence seems to point to $\mathcal{N}=4$ SYM and IIB superstring
theory being integrable models in the 't Hooft limit. We also said
that the scattering matrix is completely defined by the underlying
symmetry algebra $\mathfrak{psu}\left(2,2|4\right)$. One finds that
the $S$-matrix actually retains a symmetry algebra that is two copies
of a central extension of the $\mathfrak{psu}\left(2|2\right)$ algebra,
in particular: $\mathfrak{psu}\left(2|2\right)\ltimes\mathbb{R}^{3}=\mathfrak{su}\left(2|2\right)\ltimes\mathbb{R}^{2}.$
This symmetry of the $S$-matrix is expected to be a Yangian \cite{Beisert:2007ds,Matsumoto:2007rh,Beisert:2008tw},
and have an underlying Hopf algebra \cite{Gomez:2006va,Plefka:2006ze}
(see also \cite{Gomez:2007zr,Young:2007wd,Spill:2008tp}). Having
these new developments in minds, it would be interesting to apply
the methods used in this paper to the study of the Hopf algebra related
to the central extension, and get some results on the corresponding
Yangian generators.

The sector of near $1/2$ BPS operators in $\mathcal{N}=4$ super
Yang-Mills has been well studied by the use of collective methods
\cite{deMelloKoch:2002nq,deMelloKoch:2003pv}, and the same methods
can be used to study the elements of the algebra in $1/4$ BPS sector
(work in progress).

The author would like to thank Michael C. Abbott and Thomas Klose
for reading drafts of this work, and their very useful comments, and
also to Antal Jevicki for having helped at every step of the project.
Finally, the author would like to also thank the referee for several
very useful comments.

This work was supported in part by DOE grant DE-FG02-91ER40688- Task
A, and also supported in part by POCI 2010 and FSE, Portugal, through
the fellowship SFRH/BD/14351/2003.

\appendix

\section{Commuting the $\mathfrak{su}\left(2|2\right)$ supercharges up to
two-loops\label{sec:Commuting-the-charges-2-loops}}

The expressions found here are restrictions to the $\mathfrak{su}\left(2|2\right)$
subsector of the full sector $\mathfrak{su}\left(2|3\right)$ found
in \cite{Beisert:2003ys}. The supercharges at order $g^{0}$, $Q_{0},S_{0}$,
at order $g^{1}$, $Q_{1},S_{1}$ and at order $g^{2}$, $Q_{2},S_{2}$
in the dilute gas approximation can be written as follows:\begin{eqnarray*}
\left(Q_{0}\right)_{\beta}^{b} & = & \left\{ \begin{array}{c}
b\\
\beta\end{array}\right\} ,\\
\left(S_{0}\right)_{a}^{\alpha} & = & \left\{ \begin{array}{c}
\alpha\\
a\end{array}\right\} ;\end{eqnarray*}
 \begin{eqnarray*}
\left(Q_{1}\right)_{\beta}^{b} & = & \frac{A}{\sqrt{2}}\varepsilon_{\beta\beta'}\varepsilon^{bb'}\left(\left\{ \begin{array}{c}
\beta'\\
b'3\end{array}\right\} -\left\{ \begin{array}{c}
\beta'\\
3b'\end{array}\right\} \right),\\
\left(S_{1}\right)_{a}^{\alpha} & = & \frac{A}{\sqrt{2}}\varepsilon_{aa'}\varepsilon^{\alpha\alpha'}\left(\left\{ \begin{array}{c}
a'3\\
\alpha'\end{array}\right\} -\left\{ \begin{array}{c}
3a'\\
\alpha'\end{array}\right\} \right);\end{eqnarray*}
 \begin{eqnarray*}
\left(Q_{2}\right)_{\beta}^{b} & = & \left(\frac{A^{2}}{4}-\frac{i}{2}\gamma_{3}+\frac{i}{2}\gamma_{4}\right)\left(\left\{ \begin{array}{c}
b3\\
\beta3\end{array}\right\} +\left\{ \begin{array}{c}
3b\\
3\beta\end{array}\right\} \right)+\left(-\frac{A^{2}}{4}-i\gamma_{1}\right)\left(\left\{ \begin{array}{c}
b3\\
3\beta\end{array}\right\} +\left\{ \begin{array}{c}
3b\\
\beta3\end{array}\right\} \right),\\
\left(S_{2}\right)_{a}^{\alpha} & = & \left(\frac{A^{2}}{4}+\frac{i}{2}\gamma_{3}-\frac{i}{2}\gamma_{4}\right)\left(\left\{ \begin{array}{c}
\alpha3\\
a3\end{array}\right\} +\left\{ \begin{array}{c}
3\alpha\\
3a\end{array}\right\} \right)+\left(-\frac{A^{2}}{4}+i\gamma_{1}\right)\left(\left\{ \begin{array}{c}
\alpha3\\
3a\end{array}\right\} +\left\{ \begin{array}{c}
3\alpha\\
a3\end{array}\right\} \right).\end{eqnarray*}
 We will be using the notation used in \cite{Beisert:2003ys}. The
index $3$ above means an insertion of a field $Z$. The action of
$\left\{ \begin{array}{c}
\alpha bc\\
c\alpha b\end{array}\right\} $ on a state looks for a sequence of a fermion followed by two bosons,
and permutes them in the order 2nd boson-fermion-1st boson. As an
example in $\mathfrak{su}\left(2|3\right)$, where indices $1,2,3$
correspond to bosons and indices $4,5$ correspond to fermions, we
have \[
\left\{ \begin{array}{c}
\alpha bc\\
c\alpha b\end{array}\right\} \ket{142334452}=\ket{134234452}+\ket{242334415}.\]

Determining the anti-commutation relations, we have:\begin{eqnarray*}
\frac{2}{A^{2}}\left\{ \left(S_{1}\right)_{a}^{\alpha},\left(Q_{1}\right)_{\beta}^{b}\right\}  & = & \delta_{a}^{b}\delta_{\beta}^{\alpha}\frac{1}{A^{2}}H_{2}-\delta_{a}^{b}\left[2\left\{ \begin{array}{c}
\alpha\\
\beta\end{array}\right\} -\left\{ \begin{array}{c}
3\alpha\\
\beta3\end{array}\right\} -\left\{ \begin{array}{c}
\alpha3\\
3\beta\end{array}\right\} \right]\\
 &  & \qquad\qquad\qquad-\delta_{\beta}^{\alpha}\left[2\left\{ \begin{array}{c}
b\\
a\end{array}\right\} -\left\{ \begin{array}{c}
3b\\
a3\end{array}\right\} -\left\{ \begin{array}{c}
b3\\
3a\end{array}\right\} \right];\end{eqnarray*}
 \begin{eqnarray*}
\frac{2}{A^{2}}\left\{ \left(S_{2}\right)_{a}^{\alpha},\left(Q_{0}\right)_{\beta}^{b}\right\} +\frac{2}{A^{2}}\left\{ \left(S_{0}\right)_{a}^{\alpha},\left(Q_{2}\right)_{\beta}^{b}\right\}  & = & 2\left[\delta_{\beta}^{\alpha}\left\{ \begin{array}{c}
b\\
a\end{array}\right\} +\delta_{a}^{b}\left\{ \begin{array}{c}
\alpha\\
\beta\end{array}\right\} \right]\\
 &  & -\delta_{\beta}^{\alpha}\left[\left\{ \begin{array}{c}
3b\\
a3\end{array}\right\} +\left\{ \begin{array}{c}
b3\\
3a\end{array}\right\} \right]-\delta_{a}^{b}\left[\left\{ \begin{array}{c}
3\alpha\\
\beta3\end{array}\right\} +\left\{ \begin{array}{c}
\alpha3\\
3\beta\end{array}\right\} \right].\end{eqnarray*}
 Then the sum of these anti-commutators gives:\[
\left\{ \left(S_{1}\right)_{a}^{\alpha},\left(Q_{1}\right)_{\beta}^{b}\right\} +\left\{ \left(S_{2}\right)_{a}^{\alpha},\left(Q_{0}\right)_{\beta}^{b}\right\} +\left\{ \left(S_{0}\right)_{a}^{\alpha},\left(Q_{2}\right)_{\beta}^{b}\right\} =\frac{1}{2}\delta_{a}^{b}\delta_{\beta}^{\alpha}H_{2},\]
 where the two loop contribution for the hamiltonian (dilute gas approx)
is:\[
\frac{1}{A^{2}}H_{2}=2\left\{ \begin{array}{c}
a\\
a\end{array}\right\} +2\left\{ \begin{array}{c}
\alpha\\
\alpha\end{array}\right\} -\left\{ \begin{array}{c}
a3\\
3a\end{array}\right\} -\left\{ \begin{array}{c}
3a\\
a3\end{array}\right\} -\left\{ \begin{array}{c}
\alpha3\\
3\alpha\end{array}\right\} -\left\{ \begin{array}{c}
3\alpha\\
\alpha3\end{array}\right\} .\]

From the results presented above, we can see that we can only get
the complete order $g^{2}$ of the hamiltonian from the commutation
of the supercharges if we consider their two-loops contributions.

\section{The $Q$ Supercharges in the $\mathfrak{su}\left(2|2\right)$ sector,
on the String Side\label{sec:The-Q-charges-string}}

As can be seen in \cite{Arutyunov:2006ak,Arutyunov:2006yd}, we can
write the charges as \begin{equation}
Q_{\mathcal{M}}=\int d\sigma e^{i\alpha x_{-}}\chi\left(B_{1}\left(x,p\right)+\zeta B_{3}\left(x,p\right)+\cdots\right)+\mathcal{O}\left(\chi^{3}\right),\label{Leading order of charges}\end{equation}
 where we only kept the term linear in fermion fields, and kept all
the bosonic terms of the expansion ($B_{n}\left(x,p\right)$ is the
term with a product of $n$ bosonic fields).

The next step is to determine the Poisson brackets of two charges
with $\alpha_{1}=\alpha_{2}=1$. For example (see the appendix of
\cite{Arutyunov:2006ak})\[
\left\{ Q_{\, a}^{\alpha},Q_{\, b}^{\beta}\right\} \sim\epsilon^{\alpha\beta}\epsilon_{ab}\int_{-r}^{r}d\sigma e^{-ix_{-}}\left(x'_{-}+\frac{d}{d\sigma}f\left(x,p\right)\right),\]
 where $f\left(x,p\right)$ is a local function of the transverse
fields. The result for $\left\{ \bar{Q}_{\,\alpha}^{a},\bar{Q}_{\,\beta}^{b}\right\} $
can be obtained by conjugation. Integrating this expression, we get
\[
\left\{ Q_{\, a}^{\alpha},Q_{\, b}^{\beta}\right\} \sim\epsilon^{\alpha\beta}\epsilon_{ab}\int_{-r}^{r}d\sigma\frac{d}{d\sigma}e^{ix_{-}}=\epsilon^{\alpha\beta}\epsilon_{ab}e^{-ix_{-}\left(-r\right)}\left(e^{-i\left[x_{-}\left(r\right)-x_{-}\left(-r\right)\right]}-1\right).\]

We know that $p_{ws}=x_{-}\left(r\right)-x_{-}\left(-r\right)$. We
also impose the boundary condition $x_{-}\left(-r\right)=x_{-}^{0}$,
which is the zero mode of $x_{-}$, conjugate to $P_{+}$. Then:\begin{eqnarray*}
\left\{ Q_{\, a}^{\alpha},Q_{\, b}^{\beta}\right\}  & \sim & \frac{1}{\zeta}\epsilon^{\alpha\beta}\epsilon_{ab}e^{-ix_{-}^{0}}\left(e^{-ip_{ws}}-1\right),\\
\left\{ \bar{Q}_{\,\alpha}^{a},\bar{Q}_{\,\beta}^{b}\right\}  & \sim & \frac{1}{\zeta}\epsilon^{ab}\epsilon_{\alpha\beta}e^{ix_{-}^{0}}\left(e^{ip_{ws}}-1\right),\end{eqnarray*}
 and consequently, the central charges are $c,c^{*}$ with:\begin{equation}
c=\frac{1}{\zeta}e^{-ix_{-}^{0}}\left(1-e^{ip_{ws}}\right)e^{-ip_{ws}}.,\label{Classical central charge c}\end{equation}
 Looking at the value of the central charge here and the one obtained
from the spin-chain formalism, we can conclude that the results are
correct up to an overall phase $e^{\pm ip_{ws}}$, as long as we match
$\left\{ Q,\overline{Q}\right\} \leftrightarrow\left\{ S,Q\right\} $
and $\left\{ C,C^{\dagger}\right\} \leftrightarrow\left\{ K,P\right\} $.
This overall phase is natural, as different boundary conditions for
$x_{-}$ will differ from each other by such a phase. Also the algebra
\eqref{Commutation relations of symmetry generators} allows a $U\left(1\right)$
automorphism, which means we can always multiply all supercharges
by some phase that can depend on all central charges.

\subsubsection*{Some Comments}

In the case of $P_{+}$ infinite, the zero mode $x_{-}^{0}$ vanishes,
but the same is not true for finite light-cone momentum. This brings
some problems, as for $P_{+}$ (which is effectively the length of
the string) finite, the transverse fields don't have to vanish at
the string points, and the symmetry algebra is thus changed.

At the quantum level, both $p_{ws}$ and $x_{-}^{0}$ are promoted
to operators $\mathbf{P},\mathbf{X}_{-}^{0}$, and the central charges
are \begin{equation}
\mathbf{C}=\frac{1}{\zeta}e^{-i\mathbf{X}_{-}^{0}}\left(e^{-i\mathbf{P}}-1\right),\label{operator central charge C}\end{equation}
 and its conjugate $\mathbf{C}^{\dagger}$. $\mathbf{X}_{-}^{0}$
is the conjugate quantum operator of $\mathbf{P_{+}}$. If we consider
a state $\mathbf{P}_{+}\ket{p_{+}}=p_{+}\ket{p_{+}}$, then a state
$e^{i\alpha\mathbf{X}_{-}^{0}}\ket{p_{+}}$ obeys\begin{equation}
\mathbf{P}_{+}e^{i\alpha\mathbf{X}_{-}^{0}}\ket{p_{+}}=\left(\alpha+p_{+}\right)e^{i\alpha\mathbf{X}_{-}^{0}}\ket{p_{+}}.\end{equation}
 Because $P_{+}$ acts as the length of the string, the operator $e^{i\alpha\mathbf{X}_{-}^{0}}$
will be the length changing operator. The Hilbert space of the theory
will be a direct sum, $\mathcal{H}=\bigoplus_{p_{+}}\mathcal{H}_{p_{+}}$,
of spaces of each of the eigenvalues of $\mathbf{P}_{+}$.

\subsection{$\mathfrak{su}\left(2|2\right)$ subsector and mode expansion}

The explicit form of the charges $Q_{\mathcal{M}}$ was determined
in \cite{Arutyunov:2006ak} . The algebra $\mathcal{J}$ includes
two $\mathfrak{psu}\left(2|2\right)$ subalgebras. We will be focusing
on the $\mathfrak{psu}\left(2|2\right)_{R}$.

The leading quadratic order of (\ref{Leading order of charges}) can
be read from the results in \cite{Arutyunov:2006ak}. The fermionic
charges are at leading order:\begin{eqnarray}
Q_{\, a}^{\alpha} & = & -\frac{1}{2}\int d\sigma e^{-\frac{i}{2}x_{-}}\left[i\theta^{\alpha}\left(2P^{Y}+iY\right)_{a}+\left(2P^{Z}-iZ\right)^{\alpha}\eta_{a}^{\dagger}-\theta^{\dagger\alpha}Y'_{a}-iZ'^{\alpha}\eta_{a}+\right.\nonumber \\
 &  & \qquad\left.+\epsilon^{\alpha\beta}\epsilon_{ab}\left(i\theta_{\beta}\left(2P^{Y}+iY\right)^{b}+\left(2P^{Z}-iZ\right)_{\beta}\eta^{\dagger b}-\theta_{\beta}^{\dagger}Y'^{b}-iZ'_{\beta}\eta^{b}\right)\right]\,,\label{Supercharge left}\\
\bar{Q}_{\alpha}^{\, a} & = & \frac{1}{2}\int d\sigma e^{\frac{i}{2}x_{-}}\left[i\theta_{\alpha}^{\dagger}\left(2P^{Y}-iY\right)^{a}-\left(2P^{Z}+iZ\right)_{\alpha}\eta^{a}+\theta_{\alpha}Y'^{a}-iZ'_{\alpha}\eta^{\dagger a}+\right.\nonumber \\
 &  & \qquad\left.+\epsilon_{\alpha\beta}\epsilon^{ab}\left(i\theta^{\dagger\beta}\left(2P^{Y}-iY\right)_{b}-\left(2P^{Z}+iZ\right)^{\beta}\eta_{b}+\theta^{\beta}Y'_{b}-iZ'^{\beta}\eta_{b}^{\dagger}\right)\right]\label{Supercharge conjugate left}\\
 & = & \left(Q_{\, a}^{\alpha}\right)^{\dagger}.\nonumber \end{eqnarray}

We want to restrict ourselves to the $\mathfrak{su}\left(2|2\right)$
subsector of \cite{Beisert:2005tm}. This corresponds to keeping only
the 2 complex coordinates $Y^{a}$ and the respective conjugate momenta
$P^{y}$. These will correspond, in the SYM side, to our bosonic excitations
$\phi^{a}$, with $a=1,2$. In terms of the fermions we will be interested
in only keeping $\theta^{\alpha},\theta_{\alpha}^{\dagger}$, which
will correspond to the 2 fermionic fields $\psi_{\alpha},\psi^{\dagger\alpha}$
from SYM.

The vacuum of the fields $Z$ in Yang-Mills will, in this case, correspond
to \cite{Berenstein:2002jq}\[
\frac{1}{\sqrt{J}N^{J/2}}\mathrm{Tr}\left(Z^{J}\right)\leftrightarrow\ket{0,p^{+}}.\]

With these restrictions, the fermionic supercharges (\ref{Supercharge left})
and (\ref{Supercharge conjugate left}) become:\begin{eqnarray*}
S_{\, a}^{\alpha} & = & -\frac{1}{2}\int d\sigma e^{-\frac{i}{2}x_{-}}\left(i\theta^{\alpha}\left(2P^{Y}+iY\right)_{a}-\epsilon^{\alpha\beta}\epsilon_{ab}\theta_{\beta}^{\dagger}Y'^{b}\right),\\
Q_{\alpha}^{\, a} & = & \frac{1}{2}\int d\sigma e^{\frac{i}{2}x_{-}}\left(i\theta_{\alpha}^{\dagger}\left(2P^{Y}-iY\right)^{a}+\epsilon_{\alpha\beta}\epsilon^{ab}\theta^{\beta}Y'_{b}\right).\end{eqnarray*}

\bibliographystyle{utcaps}
\bibliography{Susy}

\end{document}